\documentstyle[12pt]{article}

\catcode`@=11
\@addtoreset{equation}{section}
\catcode`@=12
\begin{document}
\baselineskip8mm
\title{\vskip-1.7cm
The renormalization group for non-renormalizable
theories: Einstein gravity with a scalar field}
\author{A.O.Barvinsky$^{1,2\,\dag},$ \
$\!\!$A.Yu.Kamenshchik$^{2\,\dagger}$ \ $\!\!$and
I.P.Karmazin$^{2\,\dagger}$}
\date{}
\maketitle
\hspace{-0.7cm}$^{1}${\em
Theoretical Physics Institute, Department of Physics, \ University of
Alberta, Edmonton, Canada T6G 2J1}$^{*}$\\ $^{2}${\em Nuclear Safety
Institute, Russian Academy of Sciences , Bolshaya Tulskaya 52, Moscow
113191, Russia}
\begin{abstract}
\noindent We develop a renormalization-group formalism for
non-renormalizable
theories and apply it to Einstein gravity theory coupled to a scalar
field with
the Lagrangian
$L=\sqrt{g}\,[R\,U(\phi)-\frac{1}{2}\,G(\phi)\,g^{\mu\nu}\,
\partial_{\mu}\phi\,\partial_{\nu}\phi- V(\phi)]$, where
$U(\phi),\,G(\phi)$
and $V(\phi)$ are arbitrary functions of the scalar field. We
calculate the
one-loop counterterms of this theory and obtain a system of
renormalization-group equations in partial derivatives for the
functions
$U,\,G$ and $V$ playing the role of generalized charges which
substitute for
the usual charges in multicharge theories. In the limit of a large
but slowly
varying scalar field and small spacetime curvature this system gives
the
asymptotic behaviour of the generalized charges compatible with the
conventional choice of these functions in quantum cosmological
applications. It
also demonstrates in the over-Planckian domain the existence of the
Weyl-invariant phase of gravity theory asymptotically free in
gravitational and
cosmological constants.
\end{abstract}
\\
$^{*}$ Present address\\
$^{\dag}$Electronic address: barvi@page1.ualberta.ca\\
$^{\dagger}$Electronic address: grg@ibrae.ins.su\\

\section{Introduction}
\hspace{\parindent}
It is widely recognized that a consistent theory of
quantum gravity is a matter of crucial importance with regard to the
two main
challenges of modern physics: creation of a unified theory
of fundamental interactions and construction of the theory of the
quantum
origin and evolution of the Universe. The principles of the latter
theory -- quantum cosmology -- were founded many years ago by Dirac,
Wheeler and DeWitt [1] and have been further developed in recent
years when the proposals for the quantum state of the Universe
were put forward [2--4]. Among these proposals is the so-called
no-boundary prescription of Hartle and Hawking [2,3], which is
supposed
to provide a smooth transition between the quantum birth of the
Universe
and the inflationary stage of its development. The latter property is
very
important, because the theory of the inflationary expansion of the
very
early Universe [5--12] has become an integral part of the modern
cosmology.

To study physical effects of the proposed wave function of the
inflationary
Universe at a deeper level, one has to go beyond the tree-level
approximation
and, in the absence of full non-perturbative quantum field theory and
quantum
gravity, to calculate, as a first step, perturbative quantum
corrections.
This inspired the series of papers on the one-loop calculations
in quantum cosmology [13--24], which, in particular, have shown that
the
normalizability property of the cosmological wave function and the
partition
function of the inflationary universes drastically change after the
inclusion
of loop corrections [17].

However, when resorting to perturbative calculations in quantum
cosmology
we should not forget about one of the stumbling blocks on the road to
a
consistent theory of quantum gravity -- its non-renormalizability
[25,26]. It is well known that the origin of this fundamental
problem consists in the fact that the gravitational coupling constant
has a
mass dimension $-2\ (\hbar=c=1)$. Thus, the Feynman diagrams, which
contain a growing number of graviton loops, lead formally to
an infinite set of different counterterms to the gravitational
Lagrangian,
which cannot be eliminated by a renormalization procedure of the
standard type,
that is removed by the renormalization of a finite number of
parameters [27].
As regards pure gravity theory, it was shown that at the one-loop
level on mass
shell no physically relevant divergencies remain; all of them can be
absorbed
into a field renormalizations [28]. However,the pure gravity is
two-loop
non-renormalizable, even on mass shell [29]. The situation gets worse
in case
of the interaction with matter, in particular, with the scalar field.
This
theory is non-renormalizable [28] already in the  one-loop
approximation.

There are different approaches to the problem of
non-renormalizability
in quantum gravity. One can consider the Einstein gravity as a
low-energy limit
of a more general theory such as supergravity [30] or
superstring theory [31,32]. Due to the presence of the additional
symmetry,
one has a smaller number of types of divergencies in quantum
supergravity  and
can hope to find it renormalizable. As concernes superstring theory,
we can
look forward to build a finite "Theory of Everything" from it.
However, one
should recognize that the questions of the renormalizability in
supergravity
theories and finiteness of superstring ones still remain open.

Another approach to the question of renormalizability in gravity
theory is
connected with the idea of adding to the Lagrangian curvature-squared
terms
which allow one to carry out resummation of the perturbation series
and obtain
an effective renormalizable theory [33--35]. However, pursuing this
approach we
stumble upon the residues of incorrect signs at propagator poles,
which in turn
imply the problem of the breakdown of unitarity [26].

In any case, it makes sense to try to work with the usual
non-renormalizable
Einstein gravity by overcoming our fear of the infinite number of
counterterms arising in the Lagrangian as a response to an infinite
number of different types of ultraviolet divergencies. It is
interesting
that working with non-renormalizable theories we can apply
such a useful mathematical tool as renormalization-group equations.
The idea of the possibility to apply the renormalization-group
equations in the theory with a charge having negative mass dimension
was mentioned in Ref.[36]. S.Weinberg in Ref.[26] applied the concept
of the asymptotic safety to the discussion of a renormalization group
in
quantum gravity. A theory is considered to be asymptotically safe
if "essential" coupling parameters approach a fixed point as the
momentum scale of their renormalization point goes to infinity. The
condition of the asymptotic safety could be treated as a
generalization
of the notion of renormalizability, which fixes all but a finite
number of
essential coupling parameters of a theory.

The most general scheme of obtaining the renormalization-group
equations
in arbitrary non-renormalizable theories was formulated recently by
D. Kazakov [37]. In spite of the absence of the multiplicative
renormalizability, the method proposed in [37] allows one to
calculate all the
higher singularities (the poles in the dimensional regularization
scheme)
from the generalized $\beta$-functions describing the ultraviolet
divergencies without subdividing them into those related to different
parameters of the theory under consideration. However, although this
formalism
has an undoubtful theoretical significance, it can be hardly used in
the
concrete applications.

Here, we develop the renormalization-group formalism adapted for
the purposes of Einstein gravity interacting with a scalar field.
This
model seems especially interesting because it is the inflaton scalar
field
that provides the existence and subsequent termination of the de
Sitter
stage in the evolution of the Universe, which is widely reckognized
to be
responsible for the formation of the large-scale cosmological
structure
consistent with the present-day observational data.The main idea of
our
approach consists in such rearrangement of an infinite set of
counterterms in
the Lagrangian that the groups of these terms having analogous nature
are
combined together into certain functions which we shall call the
generalized
charges (in contrast to the usual charges in the traditional theory
of the
renormalization group [38,39,27,40,41]).  These generalized charges
include
implicitly all the divergencies which can appear in the theory.

We shall introduce the generalized charges as coefficients in the
expansion for
the action in powers of the curvature and in numbers of derivatives
of the
scalar field:
\footnote
{Our conventions are:$\rm{sign}\,g_{\mu\nu}=+2, g= \rm{det}
g_{\mu\nu},
R=g^{\mu\nu}R_{.\,\mu\alpha\nu}^{\alpha}\!=
g^{\mu\nu}(\partial_{\alpha}\Gamma_{\mu\nu}^{\alpha}-~\ldots),
\nabla_\mu$
is a covariant derivative with respect to $g_{\mu\nu}, \partial/
\partial x^{\alpha}=\partial_{\alpha}=,_{\alpha}$.}
        \begin{equation}
	S\,[\,g,\phi\,]=\int d^{4}x g^{1/2}
	\left\{-V(\phi)-\frac{1}{2}G(\phi)\,g^{\mu\nu}\nabla_{\mu}
	\phi\nabla_{\nu}\phi+R(g)\,U(\phi)+\ldots\right\},
	\end{equation}
where $(\ldots)$ denotes all other possible terms containing all
higher
powers of curvatures and these derivatives. It is obvious that the
action (1.1)
contains an infinite number of such structures. Assuming the
renormalization-point  independence of the bare generalized charges
(just like
as in the usual approach to the renormalization-group theory), we can
obtain an
infinite system of generalized renormalization-group equations for an
infinite set of such charges. It is certainly impossible to work with
this
system in practice. Therefore we shall have to restrict ourselves
with a finite
subsystem of generalized charges, assuming that other structures
present
in the action (1.1) are comparatively small in concrete physical
applications.
This can happen due to two different reasons. One situation allowing
us to
consider only the first few generalized charges $U(\phi), G(\phi)$
and
$V(\phi)$ is when the rest of the terms in (1.1) are negligible,
because
$\partial\phi$ and the space-time curvature are small enough to
neglect the
terms with their higher powers. Another situation corresponds to the
setting of
the physical problem with such energy scale that the running coupling
constants
of the $(\ldots)$-structures in (1.1) are negligible at this scale -
the
property called the asymptotic freedom in corresponding coupling
constants and
justifying the use of the perturbation theory.

The situation of the first type takes place in a wide class of modern
cosmological applications in the theory of the early Universe driven
by the
large inflaton scalar field, having at the inflationary stage
sufficiently
small (compared to the Planckian scale) curvature and spacetime
gradients of
the inflaton. As far as it concerns the situation of the asymptotic
freedom, it
can only follow from the properties of solutions of renormalization
group
equations and cannot be apriori used without their analysis. Anyway,
in this
paper we shall assume either of these two possibilities as a
justification for
truncating the set of generalized charges to those of eq.(1.1) and
calculating
the corresponding one-loop counterterms and $\beta$-functions. It
will turn out
that there exists a particular solution of the renormalization group
equations
demonstrating the asymptotic freedom, which can be used as an
aposteriory
argument in favour of this approach (irrespective of the
approximation of small
curvatures and field gradients).

Thus, using the first terms in action (1.1), we can find the one-loop
counterterms in a rather general form for arbitrary functions
$U(\phi),
G(\phi)$ and $V(\phi)$. They turn to be of a very complicated
non-polynomial
structure even on mass shell [28,42]. Given these one-loop
counterterms we can
construct the generalized $\beta-$functions and study a corresponding
system of
renormalization-group equations. In a certain sense this formalism
occupies an
intermediate position between the standard renormalization-group
procedure
[38,39,27,40,41] and that of Ref. [37]. The calculation of one-loop
counterterms of the theory (1.1) is a rather nontrivial problem which
can be
solved by the combination of the covariant Schwinger-DeWitt technique
[43,44]
and the background field method [45,46]. It is remarkable that
gravitational
theory coupled non-minimally to a scalar field can be simplified by
means of
the conformal transformation of the metric and scalar field
[42,47--48], so
that the nonminimal coupling between gravity and a scalar field
disappears.
Such a technique has been used in [42] for obtaining the divergent
part of the
one-loop effective action in Einstein gravity with the cosmological
term,
non-minimally coupled to the self-interacting scalar field
	\begin{equation} S\,[g,\phi]\!=\!\int d^{4}x
	g^{1/2}\!\left\{\frac{1}{k^2}(R\!-\!2\Lambda)
	-\frac{1}{2}g^{\mu\nu}\nabla_{\mu} \phi\nabla_{\nu} \phi-
	\frac{1}{2}m^{2}\phi^{2} - \frac{1}{2}\xi R\phi^{2}\!-
	\frac{\lambda}{4!}\phi^4\right\}.
	\end{equation}
The theories (1.1) and (1.2) after the corresponding conformal
transformations
have the same form
	\begin{equation}
	S\,[\,{\cal G}, \varphi\,]=\int d^{4}x\,{\cal G}^{1/2}\left
	\{\frac{1}{k^{2}}R({\cal G})- \frac{1}{2}{\cal
	G}^{\mu\nu}\bar{\nabla}_{\mu}\varphi\bar{\nabla}_{\nu}
	\varphi-\bar{V}(\varphi)\right\},
	\end{equation}
where $\cal G$ and $\varphi$ are the conformally transformed metric
and the
scalar field, $R({\cal G})\equiv\bar{R}$ - the scalar curvature with
respect to
metric \( {\cal G}_{\mu\nu} \), \(\bar{\Box\vphantom{I}}\equiv{\cal
G}^{\mu\nu}\bar{\nabla}_ \mu\bar{\nabla}_\nu \) and
$\bar{\nabla}_\mu$ - the
covariant derivative with respect to  ${\cal G}_{\mu\nu}.$
Thus, having the one-loop divergencies for the theory (1.3) expressed
in terms
of new field variables, we at the same time have the solution for the
theory
(1.1). The inverse conformal transformation gives us the needed
counterterms as
functions of original variables.

After obtaining all the counterterms we calculate the corresponding
$\beta$
- functions using our generalized formalism. The construction of the
renorma- lization-group equations is carried out in analogy with the
standard method, but, in contrast to the usual multiple-charge
renormalization
theory, this method results in the differential equations in partial
derivatives with respect to the renormalization mass parameter $t$
and the
scalar field -- the arguments of $U(\phi,t), G(\phi,t)$ and
$V(\phi,t)$. In
view of the complexity of $\beta$-functions, even the truncated set
of these
equations for generalized charges turns to be very complicated.
Moreover, these
equations  require setting the Cauchy data, and at present we don't
have
exhaustive physical principles to fix it uniquely. Therefore, instead
of a
complete rigorous setting of the boundary-value problem, we shall
study the
admissible types of the asymptotic behaviour for the generalized
charges and
compare them with the present-day phenomenological models widely used
in the
early-Universe implications.

In this way we shall study two asymptotic forms of $U(\phi), G(\phi)$
and
$V(\phi)$. The first one has a power-logarithmic dependence on the
scalar field
in the high-energy limit of large values of $\phi$. In this limit we
find a
two-parameter family of solutions for generalized charges $U(\phi),
G(\phi)$
and $V(\phi)$. It is worth noticing that the functions
$V(\phi)=\lambda\phi^4,
G(\phi)=1$ and $U(\phi)=1-\xi\phi^2/2$, usually used in
phenomenological
models, satisfy the obtained restrictions and, hence, sustain our
renormalization-group analysis. At the same time, the models without
self-interaction of the scalar field or with a minimal coupling to
gravity
are ruled out by this analysis which, thus, can serve for selecting
intrinsically consistent models. It is also interesting to note that
the
motivation for considering the non-minimally coupled scalar field
with a
self-interaction follows also from the requirement of the
normalizability of
the cosmological wave function and a reasonable probability
distribution of
inflationary cosmologies [17]. Thus the requirements of the
high-energy
(ultraviolet) quantum consistency of the theory match with the
requirements of
a reasonable dynamical scenario in the early Universe and lead to
certain
selection rules for admissible phenomenological Lagrangians.The
second
asymptotic form of generalized charges which we consider here
involves an
exponential dependence on the scalar field. These models are of
special
interest in the theory of the early Universe, because they imply a
power-law
inflation intensively discussed in the current literature [49--54].
It turns
out that these models also satisfy the consistency conditions within
the
renormalization-group approach, which impose certain relations
between the
asymptotic expressions for generalized charges and lead to their
one-parameter
family.

A remarkable property of the obtained pure power in $\phi$ solution
for the
truncated set of $U(\phi), G(\phi)$ and $V(\phi)$ is that this
solution turns
out to be {\it exact} (valid for all values of $\phi$) and describing
in the
ultraviolet limit the Weyl-invariant theory of coupled metric and the
dilaton
field, the latter beeing a purely gauge mode of the local conformal
group. This
theory turns out to be asymptotically free in the effective
renormalized
gravitational and cosmological constants, and, thus, seems to justify
in the
high-energy domain the truncation of the above type even irrespective
of the
inflationary context with small curvatures and spacetime field
gradients. The
structure of the renormalized Lagrangian of the theory shows that at
the
intermidiate energies the dynamically excited dilaton mode,
presumably, breaks
in view of its ghost nature the over-Planckian Weyl invariance and
leads to the
low-energy theory. The latter must be described by the as yet unknown
nontrivial solution of the full system of generalized renormalization
group
equations, containing the new dimensionful parameters reflecting the
broken
Weyl and scale invariances of the theory. We discuss the possible
structure of
these solutions in connection with setting the Cauchy problem for the
generalized renormalization group equations and with the low-energy
stability
of the theory.

It is worth pointing out here the relation of the above technique to
recent
work on applications of the renormalization group to quantum field
theory on
curved spacetime background (we cite here only several references
[55--63] in a
very extensive bibliography on this subject). In these papers the
gravitational
field  was basically considered at the classical level and the
problem of its
non-renormalizability did not arise: the curvature squared terms,
generated by
the renormalization procedure for the matter fields, were usually
interpreted
as a polarization of their vacuum, contributing to Einstein's
equations, but
not as the first terms in an infinite series of local interactions.
With regard
to these papers, our work can be considered as a means to justify the
truncation of this series for physical problems with slowly varying
fields and
simultaneously to find a correct (generally non-polynomial) structure
of the
first few interactions, encoded in the generalized charges of the
above type.

The paper is organized as follows. Sec.2 contains the calculation of
one-loop
divergencies for the above models. In Sec.3 we review the standard
renormalization-group method and Kazakov's formalism and apply it to
the
calculation of the generalized $\beta$-functions and the
corresponding
renormalization-group equations. In Secs.4 and 5 we analyze the
asymptotic
properties of their solutions in the context of some cosmological
models, find
the over-Planckian Weyl invariant phase of gravity theory and discuss
the
possible prospects for the further developement of this approach and
its
physical implications.

\section{One-loop divergencies in the generalized model of coupled
gravitational and scalar fields}
\subsection{Nonlinear minimally coupled scalar field}
\hspace{\parindent}
The one-loop effective action for gauge theories has a following form
in the condensed notation of DeWitt [46]:
	\begin{equation}
	i\,W_{1-{\rm loop}}=-\frac{1}{2}\,{\rm
Tr\;ln}\frac{\delta^{2}S^{\rm
	tot}[\phi]}{\delta
	\phi^{A}\delta\phi^{B}}+{\rm Tr\;ln}\,Q_{\alpha}^{\beta},
	\end{equation}
where $\varphi^{A}$ is the full set of fields, $S^{\rm
tot}[\,\phi\,\,]=S[\,\phi\,]+S_{\chi}[\,\phi\,]$ is the total action
of the
theory including the gauge-breaking term $S_{\chi}[\,\phi\,]$,
$Q_{\beta}^{\alpha}=\nabla_{\beta}^{A}(\delta\chi^{\alpha}/\delta\varphi^{A})$
is the ghost operator defined in terms of the generators of gauge
transformations of field variables $\nabla_{\alpha}^{A}$ and gauge
conditions
$\chi^{\beta}$ entering $S_{\chi}[\,\phi\,]$ and ${\rm Tr}$ is the
functional
trace.

As was mentioned in the Introduction, the first three terms of the
action (1.1)
and the action (1.2)  can be transformed into the form
	\begin{equation}
	S\,[\,{\cal G},\varphi\,]=\int d^{4}x\,{\cal
	G}^{1/2}\left\{\frac{1}{k^2}
	R({\cal G})-\frac{1}{2}{\cal
G}^{\mu\nu}\bar{\nabla}_\mu\varphi
	\bar{\nabla}_\nu\varphi-\bar{V}(\varphi)\right\}
	\end{equation}
by means of a conformal transformation of the metric and the scalar
field. So we shall carry out all the calculations for the action
(2.2)
and then use this transformation to the initial field variables in
order to get
the divergent part of the one-loop effective action for (1.1) and
(1.2).

For the calculation of the divergent part $W_{1-\rm loop}^{\rm div}$
we shall
use the background field method [45,46] and the Schwinger-DeWitt
technique
[43,44] applicable to a wide class of differential operators, to
which belong
the inverse gauge propagator $\delta^{2}S^{\rm
tot}[\phi]/\delta\phi^{A}
\delta\phi^{B}$ and ghost operator $Q^{\alpha}_{\beta}$ of the
equation (2.1).
Obtaining these operators in the background-field method looks as
follows.
We first split ${\cal G}_{\mu\nu}$ and $\varphi$ into background
fields $({\cal
G}_{\mu\nu}^{(0)},\,\varphi^{(0)})$ and quantum disturbances
$(h_{\mu\nu},\,f)$
	\begin{equation}
	{\cal G}_{\mu\nu}={\cal G}_{\mu\nu}^{(0)}+h_{\mu\nu},
	\ \varphi=\varphi^{(0)}+f.
	\end{equation}
Then, under such a splitting, we introduce a background-covariant
gauge-breaking term in the total action and expand this action in
powers of
$(h_{\mu\nu},\,f)$, so that the kernel of the quadratic term in this
expansion
will give rise to the inverse propagator $\delta^{2}S^{\rm
tot}[\phi]/\delta\phi^{A} \delta\phi^{B}$. In what follows we shall
omit the
superscript ``$^{0}$'' in the notation of the background fields
${\cal
G}_{\mu\nu}^{(0)}$ and $\varphi^{(0)}$ - the functional arguments of
the
effective action. In such notations the gauge-breaking term can be
written as
	\begin{equation}
	S_{\chi}=-\frac{1}{2k^2}\int d^{4}x\ {\cal
	G}^{1/2}{\cal G}^{\alpha\beta}
	\chi_{\alpha}\chi_{\beta},
	\end{equation}
where the background-covariant gauge conditions, which we choose
here, are the
following functions linear in quantum disturbances
	\begin{equation}
	\chi_{\alpha}\equiv\bar{\nabla}^{\mu}h_{\mu\alpha}-\frac{1}{2
}
	\bar{\nabla}_{\alpha}h,
	\end{equation}
with
	\begin{equation}
	h={\cal G}^{\mu\nu}h_{\mu\nu}
	\end{equation}
and covariant derivatives $\bar{\nabla}^{\mu}={\cal
G}^{\mu\nu}\bar{\nabla}_
{\nu}$ defined with respect to metric ${\cal G}_{\mu\nu}$.
As a result, the part of the total action  $S^{\rm tot}[\varphi]$
quadratic in
quantum perturbations can  be represented in the form
	\begin{equation}
	S^{\rm tot}_2=\frac{1}{2}\int d^{4}x\ {\cal
	G}^{1/2}\psi^A\,F_{AB}\,\psi^B.
	\end{equation}
Here $\psi^{A}\equiv (h_{\mu\nu},f)$ and the matrix
differential operator $F_{AB}(\bar{\nabla})$ is given by
	\begin{equation}
	F_{AB}(\bar{\nabla})=C_{AB}\bar{\Box\vphantom{I}}+2\Gamma_{AB
}^{\sigma}
	\bar{\nabla}_{\sigma}+W_{AB},\,\,\,\,\bar{\Box\vphantom{I}}
\equiv{\cal
	G}^{\mu\nu}\bar{\nabla}_{\mu}\bar{\nabla}_{\nu},
	\end{equation}
where the coefficient of the covariant ${\cal G}$-metric
D'Alambertian is given
by the following matrix
	\begin{eqnarray}
	&&C_{AB}=\left(\begin{array}{cc}
	\frac{1}{4k^2}C^{\mu\nu,\alpha\beta}&0\\0&1
	\end{array}\right), \\
	&&C^{\mu\nu,\sigma\rho}=\frac{1}{4}\left({\cal G}^{\mu\sigma}
	{\cal G}^{\nu\rho}+{\cal G}^{\mu\rho}{\cal G}^{\nu\sigma}-
	{\cal G}^{\mu\nu}{\cal G}^{\rho\sigma}\right).
	\end{eqnarray}

To resort to the universal algorithms of the Schwinger-DeWitt
technique, we should transform the operator (2.8) to the minimal
form. For this purpose we, first, go over to a unit matrix
coefficient of the
higher-derivative term in (2.8) by multiplying it with the matrix
$C^{AD}$
inverse to $C_{DB}$
	\begin{eqnarray}
	&&\hat{F}(\bar{\nabla})_B^A=C^{AD}F_{DB}(\bar{\nabla})=
	\bar{\Box\vphantom{I}}\hat{I}+2\hat{\Gamma}^{\sigma}
	\bar{\nabla}_{\sigma}+\hat{W}, \\
        &&C^{AD}C_{DB}=\delta_B^A.
	\end{eqnarray}
Here and in what follows we shall use an overhat to denote the matrix
acting in
the space of fields $\psi^A$ and having one contravariant and one
covariant
index:
$\hat{I}=\delta_B^A,\,\hat{\Gamma}^{\sigma}=\hat{\Gamma}^{\sigma
B}_{A},
\,\hat{W}=\hat{W}_B^A$. So the matrix unity and the matrix $C^{AD}$
above look
like
        \begin{eqnarray}
	\hat{I}=\delta_B^A=\left(\begin{array}{cc}
	\delta_{\mu\nu}^{\alpha\beta}&0\\
	0&1\end{array}\right), \\
        C^{AD}=\left(\begin{array}{cc}k^2
	C_{\alpha\beta,\mu\nu}&0\\
	0&1\end{array}\right),
	\\
	C_{\alpha\beta,\mu\nu}={\cal G}_{\alpha\mu}
	{\cal G}_{\beta\nu}+
	{\cal G}_{\alpha\nu}{\cal G}_{\beta\mu}-
	{\cal G}_{\alpha\beta}
	{\cal G}_{\mu\nu},
\label{eqn:2.15}
	\\
	\delta_{\mu\nu}^{\alpha\beta}=
	\delta_{(\mu}^{\alpha}\delta_{\nu)}^{\beta},
	\end{eqnarray}
where the indices in brackets imply their symmetrization with the
factor $1/2$.
Note that this transformation does not change the divergent part of
the
one-loop effective action, because the matrix coefficient $C_{AB}$
gives
the contribution to the effective action proportional to
$\delta^{4}(x,x)$ and
cancelled by the local measure [43].

Then we introduce a new covariant derivative
	\begin{equation}
	{\cal D}_{\mu}=\bar{\nabla}_{\mu}+\hat{\Gamma}_{\mu},
	\end{equation}
which absorbs the part of (2.11) linear in derivatives. As a result,
the
operator $\hat{F}({\cal D})$ takes the following minimal form:
	\begin{equation}
	\hat{F}({\cal D})={\cal G}^{\mu\nu}{\cal D}_
	{\mu}{\cal D}_{\nu}\hat{I}+
	\hat{P}-\frac{1}{6}\bar{R}\hat{I},
	\end{equation}
where the scalar-curvature term $-\frac{1}{6}\bar{R}\hat{I}\equiv
-\frac{1}{6}R({\cal G})\hat{I}$ has been extracted from the potential
term of
the operator for reasons of convenience.

The ghost operator $Q_{\alpha}^{\beta}$ corresponding to the
gauge-breaking
term (2.4)-(2.5) also has the form (2.18). It is defined by the gauge
transformation of the gauge (2.5) under the transformations
$\triangle^{f}h_{\mu\nu}$ of quantum disturbances
	\begin{eqnarray}
	&&h_{\mu\nu}\rightarrow h_{\mu\nu}+
	\triangle^{f}h_{\mu\nu},\,\,\,\triangle^{f}h_{\mu\nu}=
	2\bar{\nabla}_{(\mu}f_{\nu)}, \nonumber \\
	&&\chi^{\alpha}\rightarrow\chi^{\alpha}+
	Q_{\mu}^{\alpha}(\bar{\nabla})f^{\mu},
\label{eqn:2.19}
	\end{eqnarray}
(where $f^{\mu}$ is an arbitrary vector function) and reads
	\begin{equation}
	Q_{\beta}^{\alpha}(\bar{\nabla})=
	\bar{\Box\vphantom{I}}\delta_{\mu}^{\alpha}-
	\bar{R}_{\mu}^{\alpha}.
	\end{equation}

The calculation of one-loop divergences for the functional
determinants of the
minimal operators $\hat{F}({\cal D})$ (2.18) and $Q_{\mu}^{\alpha}
(\bar{\nabla})$ (2.20) can be performed by the following universal
algorithm.
Let $\hat{\tilde{F}}\equiv\tilde{F}_B^A$ be the second-order minimal
operator of the form (2.18)
	\begin{equation}
	\hat{\tilde{F}}=\tilde{\Box\vphantom{I}}\hat{I}+
	\hat{\tilde{P}}-
	\frac{1}{6}R(\tilde{g})\hat{I},
	\end{equation}
acting on some set of fields
$\varphi=\varphi^A,\,\hat{\tilde{F}}\varphi=
\tilde{F}_B^A\varphi^B,$ where $\hat{\tilde{P}}=\tilde{P}_B^A$ is an
arbitrary
matrix, $\hat{I}=\delta_B^A,
\tilde{\Box\vphantom{I}}=\tilde{g}^{\mu\nu}
\tilde{\nabla}_{\mu}\tilde{\nabla}_{\nu}$ is the
$2\omega$-dimensional
D'Alambertian and  $\tilde{\nabla}_{\mu}$ is the covariant derivative
with any
torsion-free connection covariantly conserving the metric
$\tilde{g}_{\mu\nu}.$
Let the commutator of these covariant derivatives be given by the
action of the
matrix $\hat{{\cal R}}_{\mu\nu}$:
	\begin{equation}
 	(\tilde{\nabla}_{\mu}\tilde{\nabla}_{\nu}-
	 \tilde{\nabla}_{\nu}\tilde{\nabla}_{\mu})\varphi=
	 \hat{{\cal R}}_{\mu\nu}\varphi;\,\,\, \hat{{\cal
R}}_{\mu\nu}=
	 \hat{{\cal R}}_{\mu\nu B}^{\ \ A}.
	\end{equation}
Then the logarithmically divergent part of the one-loop effective
action for
the operator (2.21) reads [44]:
	\begin{equation}
	\frac{i}{2}\,{\rm Tr\;ln}\,\tilde{F}\mid^{\rm div}=
	\frac{1}{32\pi^{2}(2-\omega)}
	\int d^{4}x\,\tilde{g}^{1/2}\,tr\,\hat{\tilde{a}}_2,
	\end{equation}
where $\omega\rightarrow 2$ is half the dimensionality of spacetime
playing the
role of the parmeter in the dimensional regularization and the DeWitt
coefficient $\hat{\tilde{a}}_2$ is defined by the
expression
	\begin{equation}
	\hat{\tilde{a}}_2=\frac{1}{180}\left\{R_{\alpha\beta\mu\nu}
	^{2}(\tilde{g})-
	R_{\mu\nu}^{2}(\tilde{g})+\tilde{\Box\vphantom{I}}
	R(\tilde{g})\right\}+
	\frac{1}{2}\hat{\tilde{P}}^2+
	\frac{1}{12}\hat{{\cal R}}_{\mu\nu}^2+
	\frac{1}{6}\tilde{\Box\vphantom{I}}\hat{\tilde{P}}.
	\end{equation}

Therefore, the divergent part $W_{1 \rm loop}^{\rm div}$ of the
one-loop
effective action (2.1) has the form
	\begin{equation}
	W_{1-\rm loop}^{\rm div}=\frac{1}{32\pi^{2}(2-\omega)}\,
	\int d^{4}x\, {\cal
	G}^{1/2}tr\,\hat{a}_2-\frac{1}{16\pi^{2}(2-\omega)} \,
	\int d^{4}x\,{\cal
	G}^{1/2}a_{2\,\mu}^{\ \mu}.
	\end{equation}
Here $\hat{a}_2$ and $a_{2\nu}^{\ \mu}$ are the DeWitt coefficients
of the
operators $\hat{F}({\cal D})$  and $Q_{\mu}^{\alpha} (\bar{\nabla})$
correspondingly. Let us first calculate the first term of this
equation.

The metric $\tilde{g}_{\mu\nu}$ and the curvatures of the algorithm
(2.24)
corresponding to the operator (2.18) are given by
	\begin{eqnarray}
	&&\tilde{g}^{\mu\nu}={\cal G}^{\mu\nu},\,
	\tilde{g}_{\mu\nu}={\cal G}_{\mu\nu}=({\cal
G}^{\mu\nu})^{-1},\\
	&&R_{\alpha\beta\mu\nu}(\tilde{g})=\bar{R}_{\alpha\beta\mu\nu
},\,
	R_{\mu\nu}(\tilde{g})=\bar{R}_{\mu\nu},\,
	R(\tilde{g})=\bar{R}
	\end{eqnarray}
while its covariant derivatives ${\cal D}_{\mu}$ and the potential
term
$\hat{P}$ can be obtained from the matrix components of the
non-minimal
operator $F_{AB}(\bar{\nabla})$ (2.8) which have the form
\begin{equation}
\Gamma_{AB}^{\sigma}=
\left(\begin{array}{cc}0&C^{\mu\nu,\lambda\sigma}\varphi_{,\lambda}\\
-C^{\alpha\beta,\lambda\sigma}\varphi_{,\lambda}&0\end{array}
\right),
\end{equation}
\begin{equation}
W_{AB}=\left(\begin{array}{cc}
C^{\mu\nu,\lambda\sigma}\,P_{\lambda\sigma}^{\alpha\beta}&
-\frac{1}{2}{\cal
G}^{\mu\nu}\,\frac{\partial\bar{V}}{\partial\varphi}\\
-C^{\alpha\beta,\lambda\sigma}\,\bar{\nabla}_{\lambda}
\bar{\nabla}_{\sigma}\varphi-\frac{1}{2}{\cal G}^{\alpha\beta}
\frac{\partial\bar{V}}{\partial\varphi}&
-\frac{\partial^2\bar{V}}{\partial\varphi^2}\end{array}\right),
\end{equation}
\begin{eqnarray}
P_{\lambda\sigma}^{\alpha\beta}&=&\frac{1}{k^2}\left\{2\,\bar{R}_
{(\lambda.\sigma).}^{\ \ \alpha\
           \ \beta}+
           2\,\delta_{(\lambda}^{(\alpha}\,\bar{R}_{\sigma)}^{\beta)}
-
           \delta_{\lambda\sigma}^{\alpha\beta}\,\bar{R}\right.
\nonumber
      \\
      &&-{\cal G}_{\lambda\sigma}\,\bar{R}^{\alpha\beta}- {\cal
           G}^{\alpha\beta}\,\bar{R}_{\lambda\sigma}+
           \frac{1}{2}\,\left.{\cal G}_{\lambda\sigma}\,{\cal
G}^{\alpha\beta}
           \,\bar{R}\right\} \nonumber
	   \\
        &&+\frac{1}{2}\,\varphi_{,\mu}\varphi_{,\nu}\,
           {\cal G}^{\mu\nu}\,\delta_{\lambda\sigma}^{\alpha\beta}-
           2\,\delta_{(\lambda}^{(\alpha}\,\varphi_{,\sigma)}
        \,\varphi^{,\beta)}+
	\bar{V}\,\delta_{\lambda\sigma}^{\alpha\beta}       \nonumber
       \\
      &&+\frac{1}{2}\,{\cal G}_{\lambda\sigma}\,
             \varphi^{,\alpha}\,\varphi^{,\beta}+\frac{1}{2}\,{\cal
G}
             ^{\alpha\beta}\,\varphi_{,\lambda}\,\varphi_{,\sigma}-
            \frac{1}{4}\,{\cal G}^{\alpha\beta}\,{\cal
G}_{\lambda\sigma}\,
         \varphi^{,\mu}\,\varphi_{,\mu}.
\end{eqnarray}
They give rise to matrices of the operator (2.11)
\begin{eqnarray}
\hat{\Gamma}^{\sigma}&=&
\mbox{}\left(\begin{array}{cc}
0&k^2\,\delta_{\rho\tau}^{\lambda\sigma}\,\varphi_{,\lambda}\\
-C^{\mu\nu,\lambda\sigma}\,\varphi_{,\lambda}&0\end{array}\right),
\\
\hat{W}&=&\left(\begin{array}{cc}
P_{\rho\tau}^{\mu\nu}&-2k^2\bar{\nabla}_{(\rho}\bar{\nabla}_{\tau)}
\,\varphi+k^2\,{\cal
G}_{\rho\tau}\,\frac{\partial\bar{V}}{\partial\varphi}
\\-\frac{1}{2}\,{\cal
G}^{\mu\nu}\,\frac{\partial\bar{V}}{\partial\varphi}&
-\frac{\partial^2\bar{V}}{\partial\varphi^2}\end{array}\right).
\end{eqnarray}
leading to the following expression for the potential term $\hat{P}$
in the
minimal form of the operator (2.18):
\begin{eqnarray}
\hat{P}&=&\hat{W}-(\bar{\nabla}_{\sigma}\,\hat{\Gamma}^{\sigma})-
{\cal G}_{\mu\nu}\,\hat{\Gamma}^{\mu}\,\hat{\Gamma}^{\nu}+
\frac{1}{6}\bar{R}\hat I  \nonumber
\\
 &=&\mbox{}\left(\begin{array}{cc}A_{\mu\nu}^{\alpha\beta}+
(1/6)\bar{R}\delta_{\mu\nu}^{\alpha\beta}&
B_{\mu\nu}\\E^{\alpha\beta}&D+(1/6)\bar{R}\end{array}\right),
\\
A_{\mu\nu}^{\alpha\beta}&=&k^2\,P_{\mu\nu}^{\alpha\beta}+
\frac{1}{2}\,k^2\delta_{(\mu}^{(\alpha}\,\varphi_{,\nu)}\,\varphi^{,\
beta)}
-\frac{1}{4}\,k^2{\cal
G}^{\alpha\beta}\,\varphi_{,\mu}\,\varphi_{,\nu},
\\
B_{\mu\nu}&=&k^2\,\left({\cal
G}_{\mu\nu}\,\frac{\partial\bar{V}}{\partial
\varphi}-\bar{\nabla}_{\mu}\,\bar{\nabla}_{\nu}\,\varphi\right),
\\
E^{\alpha\beta}&=&-\frac{1}{2}\,{\cal
G}^{\alpha\beta}\,\frac{\partial
\bar{V}}{\partial\varphi}-\frac{1}{2}\,\bar{\nabla}^{(\alpha}\,
\bar{\nabla}^{\beta)}\,\varphi+\frac{1}{4}\,{\cal G}^{\alpha\beta}\,
\bar{\Box\vphantom{I}}\varphi,
\\
D&=&-\frac{1}{2}\,\frac{\partial^2\bar{V}}{\partial\varphi^2}+
k^2\,{\cal G}^{\rho\tau}\,\varphi_{,\rho}\,\varphi_{,\tau}
\end{eqnarray}
and the corresponding commutator of covariant derivatives $({\cal
D}_{\alpha}\,
{\cal D}_{\beta}- {\cal D}_{\beta}\,{\cal
D}_{\alpha})\,\psi=\hat{\cal
R}_{\alpha\beta}\psi$:
\begin{eqnarray}
\hat{{\cal R}}_{\alpha\beta}&=&\hat{{\cal
R}}_{\alpha\beta}^0+
2\,\bar{\nabla}_{[\alpha}\,\hat{\Gamma}_{\beta]}+
2\,\hat{\Gamma}_{[\alpha}\,\hat{\Gamma}_{\beta]}     \nonumber
\\
&=&\left(\begin{array}{cc}X_{\rho\tau,\alpha\beta}^{\mu\nu}&
Y_{\rho\tau,\alpha\beta}\\
\\
Z_{\alpha\beta}^{\mu\nu}&0\end{array}\right),
\end{eqnarray}
where $\hat{{\cal R}}_{\alpha\beta}^0$ satisfies  the commutation
relation
(2.22) with the derivatives replaced by $\bar{\nabla}_\alpha$, the
square
brackets imply the antisymmetrization in indices with the factor
$1/2$ and the
blocks of the resulting matrix read
\begin{eqnarray}
X_{\rho\tau,\alpha\beta}^{\mu\nu}&=&2\delta_{(\rho}^{(\mu}
\,\bar{R}_{.\tau)\alpha\beta}^{\nu)}-
2k^2{\cal G}_{\sigma[\alpha}{\cal G}_{\beta]\kappa}\,
C^{\mu\nu,\lambda\sigma}\,\delta_{\rho\tau}^{\varepsilon\kappa}\,
\bar{\nabla}_\lambda\varphi\bar{\nabla}_{\varepsilon}\varphi,
\\
Y_{\rho\tau,\alpha\beta}&=&2k^2\,\delta_{\rho\tau}^{\lambda\sigma}\,
{\cal G}_{\sigma[\alpha}\bar{\nabla}_{\beta]}\bar{\nabla}_{\lambda}
\varphi,
\\
Z_{\alpha\beta}^{\mu\nu}&=&2\,C^{\mu\nu,\lambda\sigma}\,
{\cal G}_{\sigma[\beta}\bar{\nabla}_{\alpha]}\bar{\nabla}_{\lambda}
\varphi.
\end{eqnarray}

Using the above expressions in the algorithm (2.24), we obtain
the contribution of the first term in eq.(2.25)
\begin{eqnarray}
\left.\frac{i}{2}\,{\rm Tr\;ln}\,\hat{F}({\cal D})\,\right|^{\rm
div}\!\!\!\!\!&=\!\!\!&
\frac{1}{32\pi^{2}(2-\omega)}
\,\int d^{4}x{\cal
G}^{1/2}\,\left\{\frac{191}{180}\,\bar{R}_{\alpha\beta
\mu\nu}^{2}-\frac{551}{180}\,\bar{R}_{\alpha\beta}^2+
\frac{119}{72}\,\bar{R}^2\right. \nonumber \\
&&
+\frac{5}{4}\,k^4({\cal
G}^{\alpha\beta}\varphi_{,\alpha}\varphi_{,\beta})
^2+k^2{\cal G}^{\alpha\beta}\varphi_{,\alpha}\varphi_{,\beta}
\left(-\frac{1}{3}\,\bar{R}+k^2\bar{V}-
2\frac{\partial^2\bar{V}}{\partial\varphi^2}\right)
\nonumber
\\
&&
-\frac{13}{3}\,k^2\bar{R}\bar{V}-
\frac{1}{6}\,\bar{R}\frac{\partial^2\bar{V}}{\partial
\varphi^2}+\frac{5}{4}\,k^4\bar{V}^2-2k^2\left(\frac{\partial\bar{V}}
{\partial\varphi}\right)^2 \nonumber
\\
&&
+\frac{1}{2}\,\left(\frac{\partial^2\bar{V}}
{\partial\varphi^2}\right)^2\left.\phantom{\frac{1}{1}}\!\!\!\!\!
\right\}.
\end{eqnarray}

For the ghost operator (2.20) we have the following quantities
participating in
the algorithm (2.24) for the trace of the vector-field DeWitt
coefficient
${a_{2}}^{\mu}_{\mu}$
	\begin{eqnarray}
	&&\tilde{g}_{\mu\nu}={\cal G}_{\mu\nu},\,
	\tilde{\nabla}_\mu=\bar{\nabla}_\mu, \\
	&&\hat{\tilde{P}}=\bar{R}_{\alpha}^{\mu}+
	\frac{1}{6}\,\delta_{\alpha}^{\mu}\,\bar{R},\\
	&&\hat{{\cal R}}_{\alpha\beta}=
	\left({\cal R}_{\alpha\beta}\right)_{\mu\nu}
	^{\lambda\gamma}=-2\delta_{(\nu}^{(\gamma}
	\bar{R}_{.\,\mu)\alpha\beta}^{\lambda)}.
	\end{eqnarray}
Therefore, the ghost contribution to (2.25) equals
\begin{eqnarray}
\lefteqn{{\rm
Tr\;ln}\,[\bar{\Box\vphantom{I}}\,\delta_{\alpha}^{\mu}+
\bar{R}_{\alpha}^{\mu}]|^{\rm div}=}\nonumber \\
& &\frac{1}{16\pi^2(2-\omega)}\,\int d^{4}x\ {\cal G}^{1/2}
\left\{-\frac{11}{180}\,\bar{R}^2_{\alpha\beta\mu\nu}+
\frac{43}{90}\,\bar{R}_{\alpha\beta}^2+\frac{2}{9}\,\bar{R}^2\right\}
,
\end{eqnarray}
and the total divergent part of the one-loop effective action for the
theory
with minimally coupled nonlinear scalar field (2.2) reads
\footnote
{We used the fact that $\int d^4{x}\ {\cal G}^{1/2}
\{\bar{R}_{\alpha\beta\mu\nu}^2-4\bar{R}_{\alpha\beta}^2+\bar{R}^2\}$
is the topological invariant which can be reduced to the surface
integral.}
\begin{eqnarray}
&&W_{1-\rm loop}^{\rm div}=
\frac{1}{32\pi^2(2-w)}\int d^{4}x\ {\cal G}^{1/2}\,
\left\{\frac{43}{60}\,\bar{R}_{\alpha\beta}^2+\frac{1}{40}\bar{R}^2+
\frac{5}{4}k^4({\cal
G}^{\alpha\beta}\varphi_{,\alpha}\varphi_{,\beta}
)^2\right. \nonumber \\
&&+k^2{\cal
G}^{\alpha\beta}\varphi_{,\alpha}\varphi_{,\beta}\left(-\frac
{1}{3}\,\bar{R}+k^2\,\bar{V}-2\,\frac{\partial^2\bar{V}}{\partial
\varphi^2}\right)+\bar{R}\left(-\frac{13}{3}\,k^2\bar{V}
-\frac{1}{6}\frac{\partial^2\bar{V}}{\partial
\varphi^2}\right) \nonumber
\\
&&+\frac{5}{2}\,k^4\bar{V}^2-2\,k^2\left(\frac{\partial
\bar{V}}{\partial\varphi}\right)^2+\frac{1}{2}\,
\left(\frac{\partial^2\bar{V}}{\partial
\varphi^2}\right)^2\left.\phantom{\frac{1}{1}}\!\!\!\!\!\right\}.
\end{eqnarray}

\subsection{Reduction method of conformal transformations}
\hspace{\parindent}
The method reducing the calculations in theories (1.1) and (1.2) to
those of
(2.2) consists in the following simple observation which we shall
first
demonstrate on the example of the theory (1.2). It is a well-known
fact
[47,48]that under the following conformal transformation
	\begin{eqnarray}
	&&g_{\mu\nu}=\Omega^{-2}\,{\cal G}_{\mu\nu},\,
	\Omega^2=1+b\,\phi^2,\,b=-\frac{1}{2}\,k^2\xi, \\
	&&R=\bar{R}+6\Omega
	\bar{\Box\vphantom{I}}\Omega-12
	{\cal G}^{\mu\nu}\Omega_{,\mu}\Omega_{,\nu},\,
	\bar R\equiv R({\cal G}),
	\end{eqnarray}
the action (1.2), $S[\,g,\phi\,]$, takes in terms of the new metric
${\cal
G}_{\mu\nu}$ the following form free from the nonminimal interaction
between
the scalar and gravitational fields
	\begin{eqnarray}
	S\,[\,g,\phi\,]\!\!\!&=&\!\int d^{4}x\ {\cal
	G}^{1/2}\!\left\{\frac{1}{k^2}\bar{R}-
	\frac{1}{2}\Omega^{-4}(1-a\phi^2)\,{\cal
	G}^{\mu\nu}\phi_{,\mu}\phi_{,\nu}-W(\phi)\right\},\phantom{00
00} \\
	W(\phi)\!\!&=&\!\!\Omega^{-4}\left(\frac{2\Lambda}{k^2}+
	\frac{1}{2}m^2\phi^2+
	\frac{\lambda}{4!}\phi^4\right),
	\end{eqnarray}
where $a\equiv\frac{1}{2}k^2\xi(1-6\xi)$. However, the kinetic term
of
the Lagrangian in (2.50) contains an essential nonlinearity in
$\phi$. To
eliminate it we introduce the new scalar field $\varphi$ related to
the old one
by means of the differential equation
	\begin{equation}
	\Omega^{-4}(1-a\phi^2){\cal
	G}^{\mu\nu}\phi_{,\mu}\phi_{,\nu}= \alpha{\cal
	G}^{\mu\nu}\varphi_{,\mu}\varphi_{,\nu},
	\end{equation}
where $\alpha={\rm sign}[\Omega^{-4}(1-a\phi^4)]$. Later on we shall
confine
ourselves with $\alpha=+1$, because the negative sign corresponds to
the ghost
nature of a scalar field.

The relation (2.52) obviously holds if $\varphi=\varphi(\phi)$
satisfies the
equation
	\begin{equation}
	\left(\frac{d\varphi}{d\phi}\right)^2=\Omega^{-4}(1-a\phi^2),
	\end{equation}
which has a solution
	\begin{equation}
	\varphi=\left\{\begin{array}{l}
	-\frac{\sqrt{a}}{b}\,{\rm arcsin}\,(\sqrt{a}\phi)-
	\frac{\sqrt{-(a+b)}}{2b}\,{\rm ln}\,\frac{1+z}{1-z},\\
	\\
	z=\phi\sqrt{-(a+b)/(1-a\phi^2)},\,0\leq\xi\leq\frac{1}{6},;\\
	\\
	\\
	\frac{\sqrt{\mid a\mid}}{b}\,{\rm arcsh}\,(\sqrt{\mid
a\mid}\phi)-
	\frac{\sqrt{\mid a\mid-b}}{2b}\,{\rm ln}\,\frac{1+z}{1-z},\\
	\\
	 z=\phi\sqrt{(\mid a\mid-b)/(1-\mid a\mid\phi^2)},\,\
	 \xi<0\,\,{\rm and}\,\,\xi>1/6,
	\end{array}\right.
	\end{equation}
where the constant of integration is defined by the condition
$\varphi(\phi)\mid_{\phi=0}=0$. From this solution it is obvious
that the inverse expression $\phi=\phi(\varphi)$ cannot be obtained
analytically. This fact does not, however, present any difficulty
because for the calculation of one-loop divergencies
of the theory (1.2) (as well as for (1.1)) we shall not need an
explicit expression for the potential $\bar{V}(\varphi)$, which can
be
written formally as
	\begin{equation}
	\bar{V}(\varphi)=\left.W(\phi)\,\right|_{\phi=\phi\,(\varphi)
}.
	\end{equation}

Now the calculation of one-loop divergences in the theory (1.2)
reduces to
using the result (2.47) for a simplified model of minimal nonlinear
scalar
field (2.2) with the metric ${\cal G}_{\mu\nu}$ and the field
$\varphi$
reparametrized back to the original field variables. This
reparametrization can
be done by using the following relations
	\begin{eqnarray}
	&&
	\bar{R}=\Omega^{-2}\,R-6\,\Omega^{-6}\,b\,
	\phi_{,\alpha}\phi^{,\alpha}-
	6\,\Omega^{-4}\,b\,\phi\,\Box\phi,\\
	& &
	\int d^{4}x\,{\cal
	G}^{1/2}\,\left\{\frac{43}{60}\,\bar{R}_{\alpha\beta}^2+
	\frac{1}{40}\,\bar{R}^2\right\} \nonumber \\
	& &
	\phantom{00000000}=\int d^{4}x\
	g^{1/2}\left\{\frac{43}{60}\,R_{\alpha\beta}^2+\frac{1}{40}\,
	R^2-\frac{19}{6}\,b\,\Omega^{-4}\,R\,\phi_{,\alpha}\phi_{,
\beta}\right.
	\nonumber \\
	& &
	\phantom{00000000}-\frac{19}{6}\,b\,\Omega^{-2}\,R\,\phi\,
\Box\phi+
	\frac{19}{2}\,\Omega^{-8}\,b^2\,(\phi_{,\alpha}\phi^{,\alpha}
)^2
	\nonumber
	\\&&
	\phantom{00000000}\left.+19\,\Omega^{-6}\,b^2\,\phi_{,\alpha}
	\phi^{,\alpha}\,\phi\,\Box\phi+
	\frac{19}{2}\,\Omega^{-4}\,b^2\,\phi^2\,(\Box\phi)^2\right\},
\\
	& &
	\int d^{4}x\ {\cal G}^{1/2}\,({\cal
G}^{\alpha\beta}\,\phi_{,\alpha}
	\phi_{,\beta})^2 \nonumber \\
	& &
	\phantom{00000000}=\int d^{4}x\ g^{1/2}\,
	\Omega^{-8}(1-a\phi^2)\,(\phi_{,\alpha}
	\phi^{,\alpha})^2,\\
	& &
	\int d^{4}x\ {\cal G}^{1/2}\,\bar{R}\,{\cal
G}^{\alpha\beta}\,
	\phi_{,\alpha}\phi_{,\beta} \nonumber \\
	& &
	\phantom{00000000}=\int d^{4}x\
g^{1/2}\,(1-a\phi^2)\,\left\{\Omega^{-4}\,R\,
	\phi_{,\alpha}\phi^{,\alpha}\right. \nonumber \\
	& &
	\phantom{00000000}\left.-6\,\Omega^{-8}\,b\,
	(\phi_{,\alpha}\phi^{,\alpha})^2-6\,\Omega^{-6}\,b\,\phi_{,\alpha}
	\phi^{,\alpha}\,\phi\,\Box\phi\right\},
	\end{eqnarray}
and also the expression (2.55) for the scalar potential $\bar V$ in
the old
variables, which allows one to calculate in the same variables the
following
derivatives
\begin{eqnarray}
\frac{\partial\bar{V}}{\partial\varphi}&\equiv&
\left.\frac{\partial\bar{V}}{\partial\varphi}\right|_{\varphi=\varphi
\,(\phi)}=
\left(\frac{\partial W(\phi)}{\partial\phi}\right)
\left(\frac{\partial\varphi}{\partial\phi}\right)^{-1} \nonumber \\
&=&\Omega^{-4}\,\phi\,
\frac{(4\,b\,\lambda/k^2)+m^2+(\lambda/6)\,\phi^2-m^2\,b\,\phi^2}
{\sqrt{1-a\,\phi^2}},
\end{eqnarray}
\begin{eqnarray}
&&\frac{\partial^2\bar{V}}{\partial\varphi^2}\equiv
\left.\frac{\partial^2\bar{V}}{\partial\varphi^2}\right|_{\varphi=
\varphi\,(\phi)}=
\left(\frac{\partial^2 W(\phi)}{\partial\phi^2}\ \frac
{\partial\varphi}{\partial\phi}-
\frac{\partial W(\phi)}{\partial\phi}\ \frac
{\partial^2\varphi}{\partial\phi^2}\right)\left(\frac{\partial\varphi
}
{\partial\phi}\right)^{-3}
\nonumber \\
&&=\frac{\Omega^{-4}}{(1-a\,\phi^2)^2}\,
\left\{\frac{8\,b\,\Lambda}{k^2}+m^2+\phi^2\,(\frac{24\,b^2\,\Lambda}
{k^2}+
\frac{1}{2}\,\lambda-6\,m^2\,b)+\phi^4\,
(-\frac{32\,a\,b^2\,\Lambda}{k^2}\right. \nonumber \\
&&\left.-\frac{1}{6}\,\lambda\,b-\frac{1}{3}\,\lambda\,a+m^2\,b^2+6\,
a\,b\,m^2)
+\phi^6\,(\frac{2}{3}\,a\,b\,\lambda-2\,a\,b^2\,m^2)\right\}.
\end{eqnarray}

Finally, the one-loop divergencies of the theory (1.2) take the form:
\begin{eqnarray}
W_{1-\rm loop}^{\rm
div}[g,\phi]\!\!\!&=\!\!\!&\frac{1}{32\pi^2(2-\omega)}\,
\int d^{4}x\
g^{1/2}\,\left\{\frac{43}{60}\,R_{\alpha\beta}^2+\frac{1}{40}\,
R^2\right. \nonumber \\
& &
+\left[\frac{19}{2}\,b^2+\frac{5}{4}\,k^4\,(1-a\,\phi^2)^2+2\,k^2\,b\
,
(1-a\,\phi^2)\right]\,\Omega^{-8}\,(\phi_{,\alpha}\phi^{,\alpha})^2
\nonumber \\
& &
+\left[19\,b^2+2\,k^2\,b\,(1-a\,\phi^2)\right]\,\Omega^{-6}\,
\phi_{,\alpha}\phi^{,\alpha}\,\phi\,\Box\phi\nonumber\\
&&
+\frac{19}{2}\,\Omega^{-4}\,b^2\,\phi^2\,(\Box\phi)^2
-\frac{19}{6}\,b\,\Omega^{-2}\,R\,\phi\,\Box\phi\nonumber\\
&&
-\left[\frac{19}{6}\,b+\frac{1}{3}\,k^2\,\,(1-a\,\phi^2)\right]\,
\Omega^{-4}\,
R\,\phi_{,\alpha}\phi^{,\alpha} \nonumber \\
& &
+b\,\left(26\,k^2\,W+
\frac{\partial^2\bar{V}}{\partial\varphi^2}\right)\,\phi
\,\Box\phi+k^2\,\Omega^{-2}\,W\,\left[k^2\,(1-a\,
\phi^2)\right. \nonumber \\
& &
\left.+26\,b\right]\,\phi_{,\alpha}\phi^{,\alpha}+\Omega^{-2}\,
\frac{\partial^2\bar{V}}{\partial\varphi^2}\,
\left[-2\,k^2\,(1-a\,\phi^2)+b\right]\,\phi_{,\alpha}\phi^{,\alpha}
\nonumber
\\
& &
+\Omega^2\,\left(-\frac{13}{3}\,k^2\,W(\phi)-
\frac{1}{6}\,\frac{\partial^2\bar{V}}{\partial\varphi^2}
\right)\,R+\Omega^4\,\left[5\,k^4\,W^2(\phi)\right. \nonumber
\\
& &-2\,k^2\,\left(\frac{\partial\bar{V}}{\partial\varphi}\right)^2+
\frac{1}{2}\,\left(\frac{\partial^2\bar{V}}{\partial\varphi^2}\right)
^2
\left.\left.\!\!\!\!\!\phantom{\frac{1}{1}}\right]\right\}.
\end{eqnarray}
{}From these formulae it is clear that the one-loop divergencies
have a very complicated structure (the counterterms are
non-polynomial in the
scalar field $\phi$). Consequently, the theory with the Lagrangian
(1.2), as
well as other theories which include gravity interacting with  matter
fields,
is non-renormalizable not only due to the counterterms quadratic in
curvature,
but also because of these nonpolynomial structures.

Let us consider one limiting case for the theory (1.2) when
$\Lambda=\lambda=
m=0$. We have then $a=b=0,\ \Omega^2=1.$ Hence, the new field
variables
coincide with old ones:
\begin{equation}
{\cal G}_{\mu\nu}\equiv g_{\mu\nu},\ \varphi=\phi.
\end{equation}
The action (2.2) takes the form
\begin{equation}
S\,[\,g,\phi\,]=\int d^{4}x\
g^{1/2}\,\left\{\frac{1}{k^2}\,R-\frac{1}{2}\,
g^{\mu\nu}\,\phi_{,\mu}\phi_{,\nu}\right\},
\end{equation}
and the divergent part of the one-loop effective action is defined by
the
expression
\begin{eqnarray}
W_{1-\rm loop}^{\rm div}&=&\frac{1}{32\pi^2(2-\omega)}\,\int d^{4}x\
g^{1/2}\,
\left\{\frac{43}{60}\,R_{\alpha\beta}^2+\frac{1}{40}\,R^2\right.
\nonumber \\
& &
\left.+\frac{5}{4}\,k^4(g^{\alpha\beta}\,\phi_{,\alpha}\phi_{,\beta})
^2-
\frac{1}{3}\,k^2\,g^{\alpha\beta}\,\phi_{,\alpha}\phi_{,\beta}\,R
\right\},
\end{eqnarray}
which on mass shell, i.e. taking into account the equations of motion
\begin{equation}
\Box\phi=0,\ R=\frac{1}{2}\,k^2\,g^{\alpha\beta}\,\phi_{,\alpha}
\phi_{,\beta},
\end{equation}
coincides with the well-known result of t'Hooft and Veltman [28].
Just as
it has been expected, the expression (2.66) obtained is conformally
invariant for the case of $k^2\rightarrow\infty,\ m=0$ and $\xi=1/6$:
\begin{eqnarray}
W_{1-\rm loop}^{\rm
div}[g,\phi]\!\!\!&=\!\!\!&\frac{1}{32\pi^2(2-\omega)}\,
\int d^{4}x\
g^{1/2}\,\left\{\frac{43}{60}\,\left(R_{\alpha\beta}^2-\frac{1}{3}\,
R^2\right)\right. \nonumber \\
&+&\!\!\!\!\left.\left(\frac{19}{2}\,\phi^{-4}-\frac{79}{6}\,
\lambda\right)\left(\phi\,\Box\phi-\frac{1}{6}
\,R\,\phi^2\right)+\frac{91}{72}\,\lambda^2\,\phi^4\right\}.
\end{eqnarray}

For the generalized theory of nonminimal nonlinear scalar field (1.1)
the
analogue of the reparametrization (2.48) and (2.53) looks as follows
	\begin{eqnarray}
	&&g_{\mu\nu}=\Omega^{-2}{\cal
G}_{\mu\nu},\,\phi=\phi(\varphi),\\
	&&\Omega^2=U(\phi),\\
	&&\left(\frac{d\varphi}{d\phi}\right)^2=
	U^{-2}(\phi)\left[U(\phi)\,G(\phi)+
	3\,\left(\frac{dU}{d\phi}\right)^2\right],\\
	&&\bar{V}(\varphi)=\left.U^{-2}(\phi)\,V(\phi)\right|_{\phi=
	\phi\,(\varphi)}.
	\end{eqnarray}
It also reduces the theory to a simplified model (2.2) and allows us
to write
one of the main results of the present paper -- the following answer
for
one-loop divergences in the theory (1.1)
        \begin{eqnarray}
        W_{1-\rm loop}^{\rm
div}\!\!\!&=\!\!\!&\frac{1}{32\pi^2(2-\omega)}\,
        \int d^{4}x\
        g^{1/2}\left\{\frac{5}{2}\right.\,U^{-2}\,V^2\!\!
	-2\,U^{2}\left(\frac{\partial\bar{V}}{\partial\varphi}\right)
^2\!\!
	+\frac{1}{2}\,U^2\left(\frac{\partial^2\bar{V}}
	{\partial\varphi^2}\right)^2 \nonumber\\
        &&
        +\left[\,\left(\frac{45}{2}\,U^{-3}\,(U')^{2}+
        U^{-2}\,G\right)\,V-13\,U^{-2}\,U'\,V'\right.\nonumber\\
	&&
	-\left.\left(\frac{25}{4}\,U^{-1}\,
        (U')^{2}+2\,G+\frac{1}{2}\,
	U'\,\frac{d}{d\phi}\right)\,\frac{\partial^{2}\bar{V}}
        {\partial\varphi^{2}}\right]\,\phi_{,\mu}\phi^{,\mu}\nonumber
\\
        &&
	-\left[\frac{13}{3}\,U^{-1}\,V+\frac{1}{6}\,U\,
        \frac{\partial^2\bar{V}}{\partial\varphi^2}\right]\,R
        \nonumber \\
        &&
        +\frac{43}{60}\,R_{\alpha\beta}^2+\frac{1}{40}\,
        R^2+\frac{43}{60}\,U^{-2}\,(U^{\prime})^2
        \,R^{\alpha\beta}\,\phi_{,\alpha}
        \phi_{,\beta}-\frac{19}{12}\,
        U^{-1}\,U'\,R\,\Box\phi \nonumber \\
	&&
        -\left(\frac{5}{12}\,U^{-2}\,(U')^2
        +\frac{19}{12}\,U^{-1}U''+\frac{1}{3}\,
        U^{-1}\,G\right)\,R\,\phi_{,\mu}\phi^{,\mu} \nonumber\\
	&&
        +\left[\frac{711}{20}\,U^{-4}\,(U')^{4}+
        \frac{13}{12}\,U^{-3}\,(U')^{2}\,U''
        -\frac{113}{120}\,U^{-2}\,(U'')^{2}\right.\nonumber \\
        &&
	-\frac{199}{60}\,U^{-2}\,U'\,
        U'''+13\,U^{-3}
        (U')^{2}\,G\nonumber \\
        &&\left.-2\,U^{-2}\,U''\,G-3\,U^{-2}\,
        U'\,G'+\frac{5}{4}\,U^{-2}\,G^{2}\right]
	(\phi_{,\mu}\phi^{,\mu})^2 \nonumber \\
	&&
        -\left[\frac{26}{5}\,U^{-2}\,U'\,U''
        +\frac{111}{20}\,U^{-3}\,(U')^3+6\,U^{-2}\,U'\,G\right]\,
        \phi_{,\alpha}\phi_{,\beta}\,\phi^{;\alpha\beta} \nonumber\\
        &&
        +\left.\frac{43}{60}\,U^{-2}\,(U')^2\,
        (\phi_{;\alpha\beta})^2+\frac{199}{120}\,U^{-2}\,
        (U')^2\,(\Box\phi)^2 \right\}.
        \end{eqnarray}
Here primes are used to denote the derivatives of the generalized
charges with
respect to $\phi$
	\begin{eqnarray}
	U'=dU/d\phi,\,U''=d^2U/d\phi^2,\,U'''=d^3U/d\phi^3,\,
	G'=dG/d\phi,\,V'=dV/d\phi, \nonumber
	\end{eqnarray}
etc., and the derivatives of the potential $\bar V$ with respect to a
new
scalar field $\varphi$ are given by
        \begin{eqnarray}
        \frac{\partial\bar{V}}{\partial\varphi}&=&
        \frac{-2\,U^{-2}\,U'\,V+U^{-1}\,V'}
        {[U\,G+3\,(U')^2]^{1/2}},\\
        \frac{\partial^{2}\bar{V}}{\partial\varphi^{2}}&=&
        \frac{1}{[U\,G+3(U')^{2}]^{2}}
        \left[12\,U^{-2}\,(U')^{4}\,V
        -9U^{-1}\,(U')^{3}\,V'\phantom{\frac{1}{1}}\right.\nonumber\\
	&&
	+3\,(U')^{2}\,V''-3\,U'\,U''\,V'+
	5\,U^{-1}\,(U')^{2}\,G\,V-2\,U''\,G\,V\nonumber\\
	&&
	\left.+U\,G\,V''-\frac{7}{2}\,U'\,G\,V'+
        U'\,G'\,V-\frac{1}{2}\,U\,G'\,V'\right].
        \end{eqnarray}

It is the expression (2.72) which will be applied within the
generalized
renormalization-group approach to quantum gravity interacting with
the
scalar field.

\section{
Renormalization-group equations in non-renormalizable theories}
\subsection{Renormalization group in multicharge theories}
\hspace{\parindent}
The idea of the renormalization group theory can be expressed in
terms of bare
quantities (coupling constants, masses,fields) and counterterms to
Lagrangians.
A basic property of these bare quantities consists in the fact that
being
infinite they, after being substituted into corresponding Feynman
diagrams,
provide the cancellation of all ultraviolet divergences and give us
finite
values for all observable physical quantities, such as cross-sections
for
scattering processes, physical masses and so on. These finite charges
and
masses are called the renormalized ones, and the procedure of
eliminating the
ultraviolet divergences is called the renormalization.

The renormalization procedure requires at the intermediate stages
some
regularization which allows to avoid ill-defined quantities during
the
elimination of divergences [27]. At the final stage of calculations
one removes
the regularization and obtains finite results. However, as a remnant
of all
these operations with infinities, one gets certain ambiguity in the
final
results, which can be parametrized by a mass-dimensional parameter
$\mu$. The
origin of this parameter is different in various regularization
schemes. In the
minimal subtraction scheme of dimensional regularization [64,65],
which we
shall use in this paper, $\mu$ appears as a dimension-correcting
parameter.

Now, we can go to the definition of the renormalizability of
quantum field-theoretical models. A quantum field model is called
renormalizable if ultraviolet divergences in all orders of the
perturbation theory can be cancelled by inserting into the
Lagrangian of the theory a finite number of bare charges. In other
words, there
are only a finite number of field structures for which we
obtain divergent coefficients. All these divergences can be cancelled
by adding to the initial ``naive'' classical Lagrangian a finite
number
of counterterms. In the opposite case, when there is an infinite
number of
divergent structures, the theory is called non-renormalizable.

We have already mentioned that the renormalization-procedure
ambiguity should not affect the values of physically observable
quantities. This requirement, at least in the case of renormalizable
theories, can be rewritten as a requirement of the independence of
bare
quantities on the renormalization mass parameter $\mu^{2}$ (see
[41]). This
condition implies certain equations regulating the
dependence of renormalized charges (or other quantities) on the
renormalization mass parameter. These equations are usually called
renormalization-group equations, because different
reparametrizations of the procedure of eliminating the ultraviolet
divergences constitutes a group. Solving these renormalization
group equations in some perturbative approximation gives an
opportunity to make a partial summation of the perturbation series.

To begin with, we shall write down the renormalization group equation
for a
usual renormalizable theory with one charge (coupling constant) in
the minimal
subtraction scheme of the dimensional regularization. In this theory
a
bare charge $g_{b}$ can be expressed through a renormalized one $g$
as
\begin{equation}
g_{b}=
(\mu^2)^{\varepsilon}\left[\,g+\sum_{n=1}^{\infty}\,\frac{a_{n}(g)}
{\varepsilon^{n}}\,\right],
\end{equation}
where $\varepsilon\equiv 4-2\omega$ is a parameter of dimensional
regularization. Introducing the notion of a $\beta$-function as the
following
derivative of the renormalized charge at fixed value of the bare
charge
\begin{equation}
\left.\mu^2\frac{dg}{d\mu^2}\,\right|_{g_{\phantom{0}_{\!\!b}}}=
-\varepsilon\,g+\beta\,(g),
\end{equation}
and differentiating (3.1) with respect to $\mu^{2}$, we make the bare
charge
$g_b$ independent of $\mu^{2}$ by imposing the following equation:
\begin{equation}
0=\varepsilon\,\left[\,g+\sum_{n=1}^{\infty}\,\frac{a_{n}(g)}
{\varepsilon^{n}}\,\right]+\left(-\varepsilon\,g+\beta\,(g)\right)\,
\left[\,1+\sum_{n=1}^{\infty}\,\frac{a_{n}^{\prime}(g)}
{\varepsilon^{n}}\,\right].
\end{equation}
Then, equating the coefficients of equal powers of $\varepsilon$, we
find
\begin{equation}
\beta\,(g)=\left(g\,\frac{\partial}{\partial g}-1\right)\,a_{1}(g)
\end{equation}
and
\begin{equation}
\left(g\frac{\partial}{\partial g}-1\right)a_n(g) =\beta\,(g)\,
\frac{\partial}{\partial g}a_{n-1}(g).
\end{equation}
Thus, we see that knowing the coefficient $a_1(g)$ at the pole
$\frac{1}{\varepsilon}$, we can determine the coefficients at higher
poles
by the eq.(3.5). Besides, we see that the $\beta-$function defined by
(3.2) is
determined by the coefficient of the first-order pole in
$\varepsilon$.
The knowledge of the $\beta$-function, on the other hand, gives,
because of the
equation
\begin{equation}
\mu^2\,\frac{dg}{d\mu^2}=\beta\,(g),
\end{equation}
the dependence of $g$ on $\mu^2$, which can be transformed into the
dependence
of $g$ and relevant Green's functions on the energy scale factor.

The generalization of this method to multicharge theories is
straightforward
[66-68]. Let us suppose that we have the set of charges $g_i$ where
$i=1,\ldots,N$.
Then Eq.(3.1) and (3.2) turn into the following sets of equations
\begin{eqnarray}
&&g_{b\ i}=
(\mu^2)^{\varepsilon}\left[\,g_i+\sum_{n=1}^{\infty}\,\frac{a_{i\
n}(g_1,\ldots
,g_N)} {\varepsilon^{n}}\,\right],\\
&&\left.\mu^2\frac{dg_i}{d\mu^2}\right|_{g_{\phantom{0}_{\!\!bi}}}=
-\varepsilon\,g_i+\beta_i\,(g_1,\ldots,g_N).
\end{eqnarray}
After differentiating Eqs.(3.7) with respect to $\mu^2$ and
substituting
into the corresponding expression the definition (3.8), we compare
the
coefficients of equal powers of $\varepsilon$ and get the following
expressions for $\beta$-functions:
\begin{equation}
\beta_i(g_1,\ldots,g_N)=-a_{i\ 1}(g_1,\ldots,g_N)+
\sum_{j=1}^{N}g_j\frac{\partial a_{i\ 1}(g_1,\ldots,g_N)}{\partial
g_j}.
\end {equation}
Thus, instead of one renormalization group equation (3.6), we get the
whole
system of coupled equations:
\begin{equation}
\mu^2\,\frac{dg_i\,(\mu^2)}{d\mu^2}=\beta_i(g_1,\ldots,g_N).
\end{equation}

\subsection
{Functional charges and renormalization group equations in partial
derivatives}
\hspace{\parindent}
Although all this concerns renormalizable models, it was shown [37]
that the renormalization group equations could be generalized to
theories
with Lagrangians of arbitrary form, including non-renormalizable
ones.
The main idea consists in the assumption that the bare Lagrangian,
which can
include an infinite number of counterterms, does not depend on the
renormalization mass parameter $\mu^{2}$. This Lagrangian can be
represented in
the form
\begin{equation}
L^{b}=(\mu^2)^{\varepsilon}\,\left[L+\sum_{n=1}^{\infty}\,\frac{A_{n\
,L}}
{\varepsilon^{n}}\right],
\end{equation}
where the counterterms $A_{n\,L}$ are functionals of the renormalized
Lagrangian. Introducing the following definition of the
$\beta$-function
\[
\mu^2\,\frac{dL}{d\mu^2}=
-\varepsilon\,L+\beta_L,\]
and differentiating (3.11) with respect to $\mu^2$, we have the
following
relations:
\[\beta_L=(L\,\frac{\delta}{\delta L}-1)\,A_{1\,L},\]
\[(L\,\frac{\delta}{\delta L}-1)\,A_{n\,L}=\beta(L)\,
\frac{\delta}{\delta L}\,A_{n-1\,L}.\]
Thus, the generalized $\beta$-function of the Lagrangian is
determined by the coefficient of the first-order pole in (3.11). The
recurrent
relations give the higher-order poles, so that the only independent
function is
the coefficient of the first-order pole.

In spite of the theoretical significance of this formalism, it can
hardly be
used in the concrete theories, such as quantum gravity. Therefore our
purpose is to develop a formalism, which is less abstract than the
formalism
proposed in [37], but at the same time is convenient for treating
concrete
models, in particular, Einstein gravity theory interacting with a
scalar
field.  In the usual renormalization-group formalism we deal with
charges, masses and renormalization-field constants which are all
simple functions of $\mu^2$ - the coefficients of some special field
structures (for example, in the $\varphi^4$-model $\lambda(\mu^2)$ is
a
coefficient of $\varphi^4$, $m^2$ is a coefficient of $\varphi^2$,
etc.).
In the formalism of [37] one does not subdivide the Lagrangian
into some simpler structures.  Our formalism occupies the
intermediate
position between the two approaches mentioned above. Instead of the
usual numerical charges related to fixed field structures, we
introduce the generalized functional charges -- generally arbitrary
functions
of the scalar field, which appear in the Lagrangian as coefficients
of certain
powers of spacetime derivatives of the scalar field, powers of
spacetime
curvature and covariant derivatives of the curvature. Thus, in the
generalized
model (1.1) we consider as such charges the functions
$U(\phi),G(\phi)$ and
$V(\phi)$ which do not contain the dependence on
$\partial_{\mu}\phi$. They
enter the Lagrangian in the combinations:$R\,U(\phi),G(\phi)\partial_
{\mu}\phi\,\partial_{\nu}\phi\, g^{\mu\nu}$ and $V(\phi)$ which
contain no more
than second derivatives of field variables. It is obvious that we
must include
into the Lagrangian also terms which are quadratic in curvatures and
have a
fourth power in derivatives of a scalar field, and, generally, also
an infinite
number of different generalized charges which correspond to terms
with a
growing number of derivatives in the bare Lagrangian. As a result we
would have
an infinite system of renormalization-group equations with an
infinite number
of unknown functional variables. However we shall restrict ourselves
only to
these three terms and justify it by considering only those physical
problems
which can be characterized by intensive but slowly varying fields and
small
curvatures. As it is discussed in the next section, this
approximation makes
sense at certain stages of the early inflationary Universe.

Thus, giving up all higher-derivative terms, we truncate our system
of
renormalization group equations and reduce it to three equations with
three unknown functions. Let us deduce them. In analogy with the
usual
formalism, we introduce bare quantities
\[
U_b=(\mu^2)^{\varepsilon}\left[U+\sum_{n=1}^{\infty}
\frac{A_{n\,U}}{\varepsilon^n}\right],
\]
\begin{equation}
G_b=(\mu^2)^{\varepsilon}\left[G+\sum_{n=1}^{\infty}
\frac{A_{n\,G}}{\varepsilon^n}\right],
\end{equation}
\[
V_b=(\mu^2)^{\varepsilon}\left[V+\sum_{n=1}^{\infty}
\frac{A_{n\,V}}{\varepsilon^n}\right],
\]
where $A_{n\,U}$, $A_{n\,G}$ and $A_{n\,V}$ are the counterterms
which
correspond to structures $U$, $G$ and $V$ respectively. We should
also
define the following generalized $\beta$-functions:
\begin{eqnarray}
& &\beta_U=\mu^2\,\frac{\partial U}{\partial\mu^2}+\varepsilon\,U,\\
& &\beta_G=\mu^2\,\frac{\partial G}{\partial\mu^2}+\varepsilon\,G,\\
& &\beta_V=\mu^2\,\frac{\partial V}{\partial\mu^2}+\varepsilon\,V.
\end{eqnarray}
Now differentiating eqs.(3.12) with respect to $\mu^2$ and assuming
the
independence of the bare quantities on $\mu^2$, we have the following
equation:
\begin{eqnarray}
0&=&\mu^2\,\frac{\partial U_b}{\partial\mu^2}\nonumber \\
&=&\left[\varepsilon\,U+\mu^2\,\frac{\partial U}{\partial\mu^2}+
\sum_{n=1}^{\infty}\,\mu^2\,\frac{1}{\varepsilon^n}\,
\frac{\delta A_{n\,U}}{\delta U}\,\frac{\partial U}{\partial\mu^2}+
\sum_{n=1}^{\infty}\,\varepsilon\,\frac{A_{n\,U}}{\varepsilon^n}\right]
(\mu^2)^{\varepsilon} \nonumber \\
&+&\left[\sum_{n=1}^{\infty}\,\mu^2\,\frac{1}{\varepsilon^n}\,
\frac{\delta A_{n\,U}}{\delta G}\,\frac{\partial G}{\partial\mu^2}+
\sum_{n=1}^{\infty}\,\mu^2\,\frac{1}{\varepsilon^n}\,
\frac{\delta A_{n\,U}}{\delta V}\,\frac{\partial
V}{\partial\mu^2}\right]
(\mu^2)^{\varepsilon},
\end{eqnarray}
and corresponding equations for $G_b$ and $V_b$.
Substituting into these equations the proposed definitions
(3.13)--(3.15) for $\beta_U,\ \beta_G$ and $\beta_V$ and also
equating the coefficients of equal powers of $\varepsilon$, one can
obtain (from terms of zeroth power in $\varepsilon$)
\begin{eqnarray}
&&
\beta_U=-A_{1\,U}+\frac{\delta
A_{1\,U}}{\delta U}\,U+\frac{\delta A_{1\,U}}{\delta
G}\,G+\frac{\delta A_{1\,U}} {\delta V}\,V,\\
&&
\beta_G=-A_{1\,G}+\frac{\delta A_{1\,G}}{\delta U}\,U+
\frac{\delta A_{1\,G}}{\delta G}\,G+\frac{\delta A_{1\,G}}
{\delta V}\,V,\\
&&
\beta_V=-A_{1\,V}+\frac{\delta A_{1\,V}}{\delta U}\,U+
\frac{\delta A_{1\,V}}{\delta G}\,G+\frac{\delta A_{1\,V}}
{\delta V}\,V.
\end{eqnarray}

It should be emphasized that in contrast to usual multicharge
theories the
counterterm coefficients $A_{nU,G,V}$ are not simply functions of the
generalized charges, but their local one-point functionals (in the
sense of
parametric dependence on one point $\phi$ of the configuration space
of a
scalar field)
$A_{nU,G,V}=A_{nU,G,V}(\phi)[\,U(\phi'),G(\phi'),V(\phi')\,]$,
and therefore the multiplication of functional derivatives with the
corresponding charges in the above equations has a functional nature
which
should read as
	\begin{eqnarray}
	\frac{\delta A_{n}}{\delta U}\,U=
	\int d\phi'\;\frac{\delta A_{1\,U}
	(\phi)}{\delta U(\phi')}\;U(\phi')\nonumber
	\end{eqnarray}
(similar equations hold for the derivatives with respect to the other
charges).
However, as easily seen from eq.(2.72), $A_{nU,G,V}$ are local
functionals on
the configuration space of a scalar field, and, therefore, the
functional
derivatives of the above type represent the differential operators
with respect
to $\phi$
	\begin{eqnarray}
	\frac{\delta A_{1\,U}
	(\phi)}{\delta U(\phi')}=F(\partial/\partial\phi)\,
	\delta\,(\phi-\phi'), \nonumber
	\end{eqnarray}
etc. That is why we shall keep the condensed notation of
eqs.(3.16)-(3.19)
bearing in mind this differential-operator structure of functional
derivative
coefficients. In this way the equations (3.17)-(3.19) generalize the
relations
(3.9) for the usual $\beta$-functions in multicharge models.

Now, to calculate the generalized $\beta$-functions, we read from
$W_{1
\rm{loop}}^{\rm div}$ (2.72), which was found in the preceding
section, the
counterterms renormalizing the initial structures $V,\,G$ and $U$ in
the
Lagrangian (the negatives of the coefficients of the corresponding
structures
in (2.72)):
	\begin{eqnarray}
	A_{1\,V}&=&\frac{1}{32\pi^2}\left[\,\frac{5}{2}\,U^{-2}\,V^{2
}-
	2\,U^2
	\left(\frac{\partial\bar{V}}{\partial\varphi}\right)^{2}
	+\frac{1}{2}\,U^2\left(\frac{\partial^2\bar{V}}
	{\partial\varphi^2}\right)^{2}\right],\\
	A_{1\,G}&=&\frac{1}{32\pi^2}
	\left[\,\left(45\,U^{-3}\,(U')^2\,
	+2\,U^{-2}\,G\right)\,V-26\,U^{-2}\,U'\,V'
	\phantom{\left(\frac{1}{1}\right)^2}\right.\nonumber\\
	&&
	\phantom{000000}
	-\left.\left(\frac{25}{2}\,
	U^{-1}\,(U')^2+
	4\,G+U'\frac{d}{d\phi}\right)\left(\frac{\partial^{2}\bar{V}}
	{\partial\varphi^2}\right)\,\right],\\
	A_{1\,U}&=&\frac{1}{32\pi^2}\left[\,\frac{13}{3}\,
	U^{-1}V+
	\frac{1}{6}\,U\,\left(\frac{\partial^{2}\bar{V}}
	{\partial\varphi^2}\right)\,\right].
	\end{eqnarray}
and substitute these expressions into (3.17)--(3.19). Then, taking
into account
the eqs.(2.73)-(2.74) for $\partial\bar V/\partial\varphi$ and
$\partial^2\bar
V/\partial\varphi^2$, one can obtain the needed $\beta$-functions in
an
explicit form. The calculation of the functional-derivative terms in
(3.17)-(3.19) can be essentially simplified due to the observation
that these
terms actually represent the homogeneous transformation of all three
functional
arguments of $A_{1\,V}$, $A_{1\,G}$ and $A_{1\,U}$. From the above
expressions
it follows, however, that  $A_{1\,V}$, $A_{1\,G}$ and $A_{1\,U}$ are
homogeneous functionals of zeroth order in these arguments, and,
therefore,
these terms do not contribute to $\beta$-functions. Thus, in our
model (and in
our approximation) the $\beta$-functions reduce to the counterterms
themselves
	\begin{eqnarray}
	\beta_V=-A_{1\,V},\,\,\beta_G=-A_{1\,G},\,\,\beta_U=-A_{1\,U}
,
	\end{eqnarray}
which we present, in view of their complexity, in the Appendix as
explicit
functions of the generalized charges and their derivatives with
respect to
$\phi$.

Thus the truncated system of renormalization group equations has the
form where
we explicitly show the dependence of $\beta-$functions on generalized
charges
and their derivatives:
\begin{eqnarray}
& &\frac{\partial U}{\partial t}=\beta_U(U,U',U'',V,V',V'',G,G'),\\
& &\frac{\partial G}{\partial t}
=\beta_G(U,U',U'',U''',V,V',V'',V''',G,G',G''),\\
& &\frac{\partial V}{\partial t}=\beta_V(U,U',U'',V,V',V'',G,G').
\end{eqnarray}
In contrast to the usual renormalization-group equations, we work
with
functions which depend not only on
parameter $t$, but also have a non-trivial and unknown algebraic
dependence on
the field variables. It makes this system of equations much more
complicated
than the usual one and much more rich: instead of ordinary
differential
equations they represent the differential equations {\it in partial
derivatives}. One can say that due to the introduction of these
generalized
functional charges we rearranged our bare Lagrangian and could take
into
account infinite number of elementary terms which arise in the
process of renormalization. However, if in the usual
renormalization-group
equations, we investigate the dependence of effective charges on
$t\,=\,ln\,\mu^2$, here we have an additional problem -- the study of
the
functional structure of our generalized charges. These two
tasks: the investigation of a behaviour of generalized charges at
different
$t$ (that is a behaviour at different scales), and the investigation
of the
functional structure of these charges, are combined together in the
solution of
differential equations in partial derivatives with respect to $t$ and
$\phi$.

In view of the extremal complexity of $\beta$-functions which we
present in the
Appendix the general solution of these equations is hardly available
for
exhaustive analysis without any simplifying assumptions about the
structure of
$U(\phi),\ G(\phi)$ and $V(\phi)$. So the kind of information about
these
functions, we shall be able to extract here, will consist of the
admissible
large-field asymptotic behaviour of the generalized charges
$U(\phi),\ G(\phi)$
and $V(\phi)$, which will turn to be compatible with the conventional
choice of
these functions in numerous quantum gravitational models. This
asymptotic
behaviour unexpectedly hints us the existence of a simple {\it exact}
solution
describing the high-energy Weyl invariant and asymptotically free
phase of the
gravity theory considered in the next section.

\section{Asymptotic freedom, Weyl invariance and other implications
of the
generalized renormalization group theory}
\hspace{\parindent}
The implications of phenomenological particle-theory Lagrangians in
the theory
of the early Universe can be basically characterized by the
conditions in which
a large but slowly varying scalar field generates the effective
cosmological
constant which drives the inflationary stage of the Universe. This
means that
one mainly needs local terms in the Lagrangian of the theory in the
limit of
large $\phi$, discarding the terms of high powers in its spacetime
derivatives.
The magnitude of the corresponding spacetime curvatures in such
inflationary
models is also supposed to be much below the Planck scale. Altogether
these two
properties exactly match with the assumptions of our approximation in
the
generalized renormalization-group theory, which allowed us to
truncate the
system of equations for generalized charges. Now, to make the
formalism of
generalized renormalization group equations in partial derivatives
handlable,
we can go even further and consider only the large-field behaviour of
these
charges. For this purpose we shall look for the solution of our
system of
equations in the following asymptotic form
\begin{eqnarray}
& &U=u(t)\,\phi^{x_1}\,(ln\,\phi)^{x_2},\\
& &G=g(t)\,\phi^{y_1}\,(ln\,\phi)^{y_2},\\
& &V=v(t)\,\phi^{z_1}\,(ln\,\phi)^{z_2},
\end{eqnarray}
for $\phi\rightarrow\infty$. Note that such a combined
power-logarithmic
behaviour is natural in field theory, because the emergence of
logarithms in
the effective potentials is a well-known phenomenon underlying the
effects of
symmetry breaking and phase transitions [69].

Substituting the chosen ansatz (4.1)-(4.3) into the system
(3.23)-(3.25), we
can compare in the limit $\phi\rightarrow\infty$ the largest powers
of $\phi$
on the left- and right-hand sides of equations. This gives us the
following
two-parameter family of asymptotics (4.1)-(4.3) with arbitrary
parameters
$(x_1,x_2)$:
	\begin{equation}
	y_1=x_1-2,\ y_2=x_2,\ z_1=2\,x_1,\ z_2=2\,x_2
	\end{equation}
and the following ordinary differential renormalization-group
equations for the
coefficients $u(t)$, $g(t)$ and $v(t)$:
	\begin{eqnarray}
	\frac{du}{dt}\!\!\!&=\!\!\!&-\frac{1}{32\,\pi^{2}}\,
	\frac{13\,v}{3\,u},\\
	\frac{dg}{dt}\!\!\!&=\!\!\!&-\frac{1}{32\,\pi^2}\,
	\frac{v}{u^2}\,(2g-7x_1^2 u),\\
	\frac{dv}{dt}\!\!\!&=\!\!\!&-\frac{1}{32\,\pi^{2}}\,
	\frac{5\,v^{2}}{2\,u^{2}}.
	\end{eqnarray}
In the traditional particle-physics models and their applications in
the theory
of inflationary cosmology, the parameters of the functions
(4.1)-(4.3) in the
Lagrangian (1.2) have the following values
\footnote
{Usually the potentials describing  self-interaction of
inflaton scalar field have a more complicated polynomial structure
providing
the possibility of symmetry breaking; however, in the limit
$\phi\rightarrow
\infty$ the term $\lambda\phi^{4}$ dominates.}:
$x_1=2,\ y_1=0,\ z_1=4$, which, obviously, satisfy the obtained
restrictions
(4.4) and, therefore, do not contradict the generalized
renormalization group
theory.

The other interesting case, which can be considered, is related to
the choice of $U,\ G$ and $V$ in the exponential form. Such a choice
originates
from certain multidimensional theories which undergo dimensional
reduction
to an effective four-dimensional theory and result in a linear
combination of
exponential potentials [49]. They are interesting from the viewpoint
of
cosmological applications, because they provide the power-law
inflationary
scenario [49--54]. Thus we assume that for $\phi\rightarrow\infty$
\begin{eqnarray}
U(\phi)=u(t)\,{\rm exp}(\lambda_1\,\phi),\\
G(\phi)=g(t)\,{\rm exp}(\lambda_2\,\phi),\\
V(\phi)=v(t)\,{\rm exp}(\lambda_3\,\phi)
\end{eqnarray}
and, by using the same procedure as above, obtain the following
relations for
$\lambda_{i}$:
\begin{equation}
\lambda_2=\lambda_1,\ \lambda_3=2\,\lambda_1,
\end{equation}
similar to restrictions (4.4) and forming another one-parameter
family of
high-energy asymptotics.

It is interesting that the above mentioned homogenety of the one-loop
counterterms in the generalized charges, which allowed us easily to
calculate
the $\beta$-functions (3.23), and the homogenety properties of
$\beta$-functions themselves give a one-parameter family of the {\it
exact}
solutions (4.1)-(4.3) with $x_2=y_2=z_2=0$ and the coefficients
$u(t)$, $g(t)$
and $v(t)$ satisfying the set of {\it exact} equations (4.5)-(4.7). A
trivial
integration of the latter then gives the particular family of exact
generalized
charges:
	\begin{eqnarray}
	&&V(\phi,\,t)=-\frac{192}{37}\,\pi^2
C^2\,t^{15/37}\,\phi^{2x_1},\\
	&&G(\phi,\,t)=-\left(3x^2 C\,t^{26/37}-
	C_1\,t^{12/37}\right)\,\phi^{x_1-2},\\
	&&U(\phi,\,t)=C\,t^{26/37}\,\phi^{x_1},
	\end{eqnarray}
where $C$ and $C_1$ are two integration constants. Apart from
negative overall
coefficients in (4.12) and (4.13) this solution corresponds to the
asymptotic
freedom in the high-energy limit, because the effective gravitational
constant
$1/U$ vanishes in this limit, $t\rightarrow\infty$, and the growth of
the
non-linear scalar potential (4.12) is compensated by the even faster
growth of
the coefficient of the scalar kinetic term $G$ (4.13) (which means
that the
contribution of the higher-order Feynman graphs with scalar loops
will be
highly suppressed by the powers of a scalar field propagator
proportional to
$1/G$). The negative sign in (4.12), however, means that the scalar
potential
is negative, which apparently corresponds to the well-known property
of the
pure $\lambda\phi^4$-theory beeing asymptotically free only for the
wrong sign
of $\lambda$. This makes the only fixed point $2g=7x^2u$ of eq.(4.6)
unstable
and, moreover, implies in view of eqs.(4.13)-(4.14) that the
effective
gravitational constant and the kinetic term of the scalar field are
of opposite
signs, whence either the graviton or the scalar boson are supposed to
be a
ghost particle.

A possible qualitative interpretation of this seemingly unreasonable
solution
might consist in the following observation. Note that in the
high-energy limit
$t\rightarrow\infty$ there holds a relation
$G(\phi,\,t)=-3x_1^2\,U(\phi,\,t)
/\phi^2$ between the asymptotic behaviours of the generalized charges
(4.13)
and (4.14). By redefining the old scalar field from $\phi$ to a new
one
$\varphi=\phi^{x_1/2}$ one can use this relation to show that the
renormalized
action (1.1) in this limit takes the form
	\begin{eqnarray}
	S\,[\,g,\phi\,]=\!\int d^{4}x g^{1/2}
	\left\{\!\phantom{\frac{1}{1}}\!\!\!\!\!\!\right.\!\!\!&-
	\!\!\!&\frac{192}{37}\,\pi^2\,C^2\,t^{15/37}\varphi^4
	+C\,t^{26/37}\left[R(g)\,\varphi^2+
	6\,(\nabla\varphi)^2\right]\nonumber\\
	&+\!\!\!&\left.\phantom{\frac{1}{1}}\!\!\!\!\!
	O\,(\,t^{12/37})\,\right\},\,\,\;\;\varphi=\phi^{x_1/2},\,\,\
;\;
	t\rightarrow\infty,
	\end{eqnarray}
where $O\,(\,t^{12/37})$ includes both the second term of eq.(4.13)
and the
$O\,(\,t^0)$ terms of eq.(2.72) of higher powers in spacetime
derivatives and
curvatures. But this action is conformally invariant under the local
Weyl
transformations of the metric and scalar field
	\begin{equation}
	g_{\mu\nu}^{\;\prime}=g_{\mu\nu}\,\Omega^2,\;\;\;
	\varphi\,'=\varphi\,\Omega^{-1},
	\end{equation}
whence it follows that a particular (monomial in $\phi$) solution
(4.12)-(4.14)
of our renormalization group equations describes in the ultraviolet
limit a
conformally-invariant phase of the gravity theory
\footnote
{A similar high-energy Weyl invariance of the nonminimal coupling
between the
scalar field and spacetime curvature was found in refs.[57,60,62,63]
in the
context of the renormalization group theory for quantized matter in
the
external gravitational field.}.
The wrong sign of the kinetic term of the field $\varphi$ and its
quartic
interaction in (4.15) does not mean the physical instability of the
theory,
because this field is unphysical and represents a purely gauge mode
of local
transformations (4.16), which can (and must be) be gauged away by
either
imposing the conformal gauge condition $\varphi=1$ or absorbing this
field into
the redefinition of the metric field
	\begin{equation}
	{\cal
G}_{\mu\nu}=g_{\mu\nu}\,\varphi^2=g_{\mu\nu}\,\phi^{x_1}.
	\end{equation}
In terms of this new metric the action (4.15) takes the form
	\begin{eqnarray}
	S\,[\,g,\phi\,]\!=\!\int\!d^{4}x\,{\cal G}^{1/2}\!
	\left\{C\,t^{26/37}\!\left[R({\cal G})-
	\frac{192}{37}\,\pi^2\,C\,t^{-11/37}\right]+
	O\,(\,t^{12/37})\right\}
	\end{eqnarray}
of the asymptotically free Enstein theory with the positive (for
$C>0$)
gravitational and cosmological constants
	\begin{equation}
	k^2=C^{-1}t^{-26/37},\;\;\;\Lambda=
	\frac{192}{37}\,\pi^2\,C\,t^{-11/37},
	\end{equation}
both vanishing in the high-energy limit and providing the smallness
of higher
order quantum perturbation corrections.

At lower energy scales the theory (4.15) looses its Weyl invariance,
the scalar
field $\varphi$ (or $\phi$) becomes dynamical and due to its ghost
nature
induces the instability of the Weyl (and scale) invariant phase.
Therefore, for
intermidiate energies this rules out the above simple exact solutions
of
monomial type in $\phi$. In full accordance with the loss of
conformal
invariance, the possible alternative solutions of our renormalization
group
equations will have a polynomial structure in $\phi^{x_1}$ which
necessarily
induces in the theory the extra dimensional scale -- the dimensional
coefficients of different powers of $\phi$ (note that in the above
formalism of
monomial functional charges the scalar field and the constant $C$
were subject
to only one dimensional restriction: the gravitational constant
$1/U\sim1/C\phi^{x_1}$ had to be of the squared length
dimensionality). But
these dimensionful quantities can enter the theory with generalized
functional
charges only through certain dimensional parameters of the quantum
state of the
theory, such as the value of the scalar field in a stable vacuum of
the theory
with broken symmetry. These parameters can constitute at least a part
of the
full initial data in the Cauchy problem for our renormalization group
equations, which is supposed to select a unique solution for the
generalized
functional charges. Presumably, this Cauchy problem has to be posed
at some
intermidiate or low energy scale which corresponds to the low-energy
physics of
the observable Universe described by excitations over some stable
vacuum state
with broken conformal and scale invariance. Such a stable vacuum
state and its
dimensionful parameters are, in their turn, usually determined from
the
condition of stationarity of the corresponding effective potential
(or more
generally of the full effective action) with respect to the mean
scalar and
other fields. Thus the Cauchy problem for the generalized
renormalization group
equations is related to the effective equations selecting a stable
quantum
state. This approach might represent a plausible developement of the
proposed
formalism in application to the above model of quantum gravity
theory, but it
goes beyond the scope of this paper.

\section{Discussion and conclusions}
\hspace{\parindent}
Unfortunately, at the moment the picture, obtained thus far, does not
imply
much predictive power at intermidiate energy scales and does not give
the
quantitative mechanism of transition between the possible high-energy
Weyl
invariant phase of the theory and our low-energy realm. One should
bear in mind
that the above interpretation has basically a qualitative nature,
because at
present we don't have a rigorous formalism incorporating the
dynamical
transition of the theory "defreezing" purely gauge modes into the
physical
dynamical ones. This problem is analogous to the issue of a rigorous
quantization of classical gauge modes acquiring the dynamical content
at the
quantum level due to anomalies, which now has a well-established
status only in
simple low-dimensinal field theories
\footnote
{In connection with this one should mention an interesting approach
extending
the methods of the two-dimensional string models to the quantization
of the
conformal factor in 4-dimensional gravity theory, undertaken in
[70,71]. These
references also contain the renormalization-group construction of a
stable
conformally invariant phase in the infrared limit of gravity theory
-- the
domain which might be also attained within our approach via the as
yet unknown
solutions of the generalized renormalization group equations.
}.
Nevertheless, our approach would seem to give certain selection rules
for the
admissible Lagrangians of the inflaton scalar field nonminimally
coupled to
gravity and predict the existence of its nontrivial high-energy phase
with very
attractive features of asymptotic freedom and Weyl invariance. Just
to
summarize the difficulties of the above model and of the whole
formalism, let
us briefly consider the questions of principal arising in the
proposed
generalized renormalization group technique, which can serve as a
guiding
principle for the possible further developement of this approach.

The fundamental problem, which remains beyond the reach of our
considerations,
is the setting of the boundary-value problem for the renormalization
group
equations (3.23)-(3.25). Since the $\beta$-functions on their
right-hand sides
involve the generalized charges -- functions of $t$ and $\phi$ -- as
well as
their derivatives with respect to $\phi$, these boundary conditions
consist in
the Cauchy data, that is the functions of $\phi$ at some "moment" of
$t$. These
functions replace the initial values of usual charges in multi-charge
theories
at some fixed energy scale. To see it, notice that our generalized
charges are
actually the result of partial summation in the theory with an
infinite number
of usual (numerical) charges: the expansion of $U(\phi),\ G(\phi)$
and
$V(\phi)$ in powers of $\phi$ recovers the infinite set of these
usual charges
as coefficients of this expansion. Therefore, the infinity of their
initial
values can be encoded in the functions of $\phi$, which comprise the
initial
data for our renormalization group equations. Unfortunately, we don't
have at
present exhaustive physical principles to fix this data, except the
considerations, briefly mentioned above and relating this Cauchy
problem to the
search for stable quantum states of the theory.

Another approach to these equations, actually dominating the
renormalization
group theory, consists in the analyses of the fixed points of
(3.23)-(3.25) and
does not essentially require the knowledge of this initial data.
Again, a
nontrivial generalization of the usual equations for fixed points,
\begin{equation}
\beta_U=0,\ \beta_G=0,\ \beta_V=0,
\end{equation}
is that in our case these equations are not algebraic, but rather
represent
ordinary differential equations of high order in derivatives with
respect to
$\phi$. The analyses of these equations, which goes beyond the scope
of this
paper, would give the answers to the problem of the high-energy
behaviour of
this conventionally non-renormalizable theory, the ultraviolet or
infrared
stability of the fixed points, structure of the renormalization-group
flows,
etc. These equations will also require the constants of integration
(the Cauchy
problem of lower functional dimensionality) which again might be read
off the
stable quantum states in the theory.

This analysis would raise the basic conceptual issue behind the
approximate
nature of our generalized renormalization-group approach -- the
justification
for the truncation of the system of charges to a finite set of the
first few
ones $U(\phi),\ G(\phi)$, $V(\phi)$, etc. Our truncation was based on
the
physical assumption that in concrete problems under consideration the
contribution of higher-order charges is negligible because of a
slowly varying
nature of the scalar field and small curvatures of spacetime. This
assumption
can at best be justified only at the heuristic level, for virtual
quantum
disturbances of fields always probe in the renormalized Lagrangian
arbitrarily
high powers of their derivatives. The fundamental solution of this
problem
would consist in the formulation of the generalized condition of
asymptotic
freedom or safety, which basically would reduce to a statement that
at fixed
points of the generalized renormalization group flows all
higher-order
functional charges go to zero and thus justify our approximation.
Anyway, the
approach of this paper and a particular model demonstrating its
complexity
raise more questions than physically sensible predictions, but,
probably, pave
a path to more constructive attempts to renormalize conventionally
non-renormalizable theories.\\

{\bf ACKNOWLEDGEMENTS}

The authors benefitted from helpful discussions with I.L.Buchbinder
and
D.I.Kazakov and are also grateful to Don.N.Page for his help in the
preparation
of this paper for publication. One of the authors (A.O.B.) is
grateful for the
partial support of this work provided by CITA National Fellowship.
\\

{\bf APPENDIX}

We list here the expressions for the $\beta$-functions (3.23), with
due regard
for equations (3.20)--(3.22) and (2.73), (2.74) for the counterterms:
\begin{eqnarray*}
\beta_{V}\!\!\!\!&=\!\!\!\!&-\frac{1}{32\pi^{2}\,[U\,G+3(U')^{2}]^{4}
}\,
\left\{\frac{117}{2}\,U^{-2}\,(U')^{8}\,V^{2}+
108\,U^{-1}\,(U')^{7}\,V\,V'\right.\\
&&-\frac{27}{2}\,(U')^{6}\,(V')^{2}+\frac{9}{2}\,U^{2}\,(U')^{2}\,
(U'')^{2}\,(V')^{2}+\frac{9}{2}\,U^{2}\,(U')^{4}\,(V'')^{2}\\
&&-36\,(U')^{5}\,U''\,V\,V'
+36\,(U')^{6}\,V\,V''-9\,U^{2}\,(U')^{3}\,U''\,V'\,V''\\
&&+27\,U\,(U')^{4}\,U''\,(V')^{2}-27\,U\,(U')^{5}\,V'\,V''\\
&&+G\,\left(114\,U^{-1}\,(U')^{6}\,V^{2}-\frac{45}{2}\,U\,(U')^{4}\,
(V')^{2}+129\,(U')^{5}\,V\,V'\right.\\
&&-24(U')^{4}\,U''\,V^{2}+27\,U\,(U')^{4}\,V\,V''+3\,U\,(U')^{3}\,U''
\,V\,V'\\
&&-\frac{39}{2}\,U^{2}\,(U')^{3}\,V'\,V''
+6\,U^{2}\,U'\,(U'')^{2}\,V\,V'+\frac{21}{2}\,U^{2}\,(U')^{2}\,U''\,(
V')^{2}\\
&&-3\,U^{3}\,U'\,U''\,V'\,V''-6\left.\,U^{2}\,(U')^{2}\,U''\,V\,V''
+3\,U^{3}\,(U')^{2}\,(V'')^{2}\right)\\
&&+G'\left(-3\,U^{2}\,(U')^{2}\,U''\,V\,V'+
\frac{3}{2}\,U^{3}\,U'\,U''\,(V')^{2}+3\,U^{2}\,(U')^{3}\,V\,V''
\right.\\
&&-\left.\frac{3}{2}\,U^{3}\,(U')^{2}\,V'\,V''+12\,(U')^{5}\,V^{2}
-15\,U\,(U')^{4}\,V\,V'+\frac{9}{2}\,U^{2}\,(U')^{3}\,(V')^{2}\right)
\\
&&+G^2\left(\frac{151}{2}\,(U')^{4}\,\,V^{2}
-\frac{95}{8}\,U^{2}\,(U')^{2}\,(V')^{2}
+\frac{109}{2}\,U\,(U')^{3}\,V\,V'\right.\\
&&+2\,U^{2}\,(U'')^{2}\,V^{2}
+\frac{1}{2}\,U^{4}\,(V'')^{2}-10\,U\,(U')^{2}\,U''\,V^{2}\\
&&+5\,U^{2}\,(U')^{2}\,V\,V''-2\,U^{3}\,U''\,V\,V''
-\frac{7}{2}\,U^{3}\,U'\,V'\,V''+\left.7\,U^{2}\,U'\,U''\,V\,V'\right
)\\
&&+(G')^{2}\left(\frac{1}{2}\,U^{2}\,(U')^{2}\,V^{2}
+\frac{1}{8}\,U^{4}\,(V')^{2}-\frac{1}{2}\,U^{3}\,U'\,V\,V'\right)\\
&&+G\,G'\left(5\,U\,(U')^{3}\,V^{2}-6\,U^{2}\,(U')^{2}\,V\,V'
-2\,U^{2}\,U'\,U''\,V^{2}\right.\\
&&+\left.U^{3}\,U''\,V\,V'+\frac{7}{4}\,U^{3}\,U'\,(V')^{2}
+U^{3}\,U'\,V\,V''-\frac{1}{2}\,U^{4}\,V'\,V''\right)\\
&&+G^3\left(22\,U\,(U')^{2}\,V^{2}+8\,U^{2}\,U'\,V\,V'
-2\,U^{3}\,(V')^{2}\right)+\left.\frac{5}{2}\,U^{2}\,G^{4}\,V^{2}
\right\},\\
\\
\beta_{G}\!\!\!\!&=\!\!\!\!&-\frac{1}{32\pi^{2}\,[U\,G+3(U')^{2}]^{3}
}
\left\{837\,U^{-3}\,(U')^{8}\,V-\frac{855}{2}\,U^{-2}\,(U')^{7}\,V'
\right.\\
&&-\frac{171}{2}\,U^{-1}\,(U')^{6}\,V''
+\frac{171}{2}\,U^{-1}\,(U')^{5}\,U''\,V+27\,(U')^{4}\,U''\,V''\\
&&-27\,(U')^{3}\,(U'')^{2}\,V'+9\,(U')^{4}\,U'''\,V'-9\,(U')^{5}\,
V'''\\
&&+G\left(-\frac{1617}{4}\,U^{-1}\,(U')^{5}\,V'+
\frac{1701}{2}\,U^{-2}\,(U')^{6}\,V
+57\,U^{-1}\,(U')^{4}\,U''\,V\right.\\
&&-\frac{177}{2}\,(U')^{4}\,V''-24\,(U')^{2}\,(U'')^{2}\,V
+69\,(U')^{3}\,U''\,V'\\
&&+9\,U\,(U')^{2}\,U''\,V''+3\,U\,U'\,(U'')^{2}\,V'+3\,U\,(U')^{2}\,U'''
\,V'\\
&&\left.-6\,U\,(U')^{3}\,G\,V'''+6\,(U')^{3}\,U'''\,G\,
V\phantom{\frac{1}{1}}\!\!\!\!\right)\\
&&+G'\left(-\frac{57}{2}\,U^{-1}\,(U')^{5}\,V
+\frac{39}{4}\,(U')^{4}\,V'+15\,(U')^{3}\,U''\,V
-12\,U\,(U')^{2}\,U''\,V'\right.\\
&&\left.+\frac{9}{2}\,U\,(U')^{3}\,V''\right)
+G''\left(-3\,(U')^{4}\,V+\frac{3}{2}\,U\,(U')^{3}\,V'\right)\\
&&+G^2\left(\frac{607}{2}\,U^{-1}\,(U')^{4}\,V+35\,(U')^{2}\,U''\,V
-\frac{497}{4}\,(U')^{3}\,V'-32\,U\,(U')^{2}\,V''\right.\\
&&+\left.\frac{35}{2}\,U\,U'\,U''\,V'+2\,U\,U'\,U'''\,V
-U^{2}\,U'\,V'''\right)\\
&&+G\,G'\left(-\frac{35}{2}\,(U')^{3}\,V-3\,U\,U'\,U''\,V
+\frac{29}{4}\,U\,(U')^{2}\,V'+\frac{3}{2}\,U^{2}\,U'\,V''\right)\\
&&+(G')^2\left(2\,U\,(U')^{2}\,V-U^{2}\,U'\,V'\right)
+G\,G''\left(-U\,(U')^{2}\,V+\frac{1}{2}\,U^{2}\,U'\,V'\right)\\
&&+G^3\left(43\,(U')^{2}\,V-12\,U\,U'\,V'+8\,U\,U''\,V
-4\,U^{2}\,V''\right)\\
&&\left.+G^{2}\,G'\left(-4\,U\,U'\,V+2\,U^{2}\,V'\right)
+2\,U\,G^{4}\,V\right\},\\
\\
\beta_{U}\!\!\!\!&=\!\!\!\!&-\frac{1}{32\pi^{2}\,[U\,G+3(U')^{2}]^{2}
}
\left\{41\,U^{-1}\,
(U')^{4}\,V-\frac{3}{2}\,(U')^{3}\,V'\right.-\frac{1}{2}\,U\,U'\,U''
\,V'\\
&&+\frac{1}{2}\,U\,(U')^{2}\,V''\\
&&+G\left(\frac{161}{6}\,(U')^{2}\,V-\frac{1}{3}\,U\,U''\,V
-\frac{7}{12}\,U\,U'\,V'+\frac{1}{6}\,U^{2}\,V''\right)\\
&&+G'\left(\frac{1}{6}\,U\,U'\,V-\frac{1}{12}\,U^{2}\,V'\right)
\left.+\frac{13}{3}\,U\,G^{2}\,V\right\}.
\end{eqnarray*}
\\

{\bf REFERENCES}\\
\vspace{-0.6cm}
\baselineskip6mm
\begin{description}
\item[{[\rm 1]}] B.S. DeWitt, Phys. Rev. {\bf{160}}, 1113 (1967).
\item[{[\rm 2]}] J.B. Hartle and S.W. Hawking, Phys. Rev. D
 {\bf{28}}, 2960 (1983).
\item[{[\rm 3]}] S.W. Hawking,
Nucl. Phys. B {\bf{239}}, 257 (1984).
\item[{[\rm 4]}] A. Vilenkin, Phys. Rev. D {\bf{30}}, 509
(1984); {\bf{33}}, 3560 (1986).
\item[{[\rm 5]}] A.A. Starobinsky, Pisma ZhETF(USSR) {\bf 30},
 719 (1979).
\item[{[\rm 6]}] A.A. Starobinsky, Phys.Lett. {\bf{91B}}, 99
(1980).
\item[{[\rm 7]}] A.H. Guth, Phys. Rev. D {\bf{23}}, 347 (1981).
\item[{[\rm 8]}] K. Sato, Mon. Not. R. Astron. Soc. {\bf{195}},
 467 (1981).
\item[{[\rm 9]}] K. Sato, Phys. Lett. {\bf{99B}},
66 (1981).
\item[{[\rm 10]}] A.D. Linde, Phys. Lett. {\bf{108B}}, 389 (1982).
\item[{[\rm 11]}] A. Albrecht and P.J. Steinhardt,
 Phys. Rev. Lett. {\bf{48}}, 1220 (1982).
\item[{[\rm 12]}] A.D. Linde, Phys. Lett. {\bf{129B}}, 177 (1983).
\item[{[\rm 13]}] K. Schleich, Phys. Rev. D {\bf 32}, 1889 (1985).
\item[{[\rm 14]}] J. Louko, Phys. Rev. D {\bf 38}, 478 (1988).
\item[{[\rm 15]}] J. Louko, Ann. Phys. (N. Y.) {\bf 181}, 318 (1988).
\item[{[\rm 16]}] P.A. Griffin and D.A. Kosower, Phys. Lett. B {\bf
233}, 295 (1989).
\item[{[\rm 17]}] A.O. Barvinsky and
A.Yu. Kamenshchik, Class. Quantum Grav. {\bf 7}, L181 (1990).
\item[{[\rm 18]}] I. Moss and S. Poletti, Nucl. Phys. B {\bf 341},
 155 (1990).
\item[{[\rm 19]}] I. Moss and S. Poletti, Phys. Lett. B ,{\bf
245}, 355 (1990).
\item[{[\rm 20]}] P.D. D'Eath and G.V.M. Esposito,
Phys.Rev.{\bf D43} (1991) 3234; {\bf D44} (1991) 1713.
\item[{[\rm 21]}] A.O. Barvinsky, A.Yu. Kamenshchik, I.P. Karmazin
and I.V. Mishakov, Class. Quantum Grav. {\bf 9}, L27 (1992).
\item[{[\rm 22]}] A.Yu. Kamenshchik and I.V. Mishakov,
Int. J. Mod. Phys. A {\bf 7}, 3713 (1992).
\item[{[\rm 23]}] A.O. Barvinsky, A.Yu. Kamenshchik and I.P.
Karmazin, Ann. Phys. (N. Y.) {\bf 219},201 (1992).
\item[{[\rm 24]}] A.Yu. Kamenshchik and I.V. Mishakov, Phys. Rev. D
{\bf 47},
1380 (1993).
\item[{[\rm 25]}] N.D.Birrell and
P.C.W.Davies, {\it Quantum Fields in Curved Space} (Cambridge
University Press, Cambridge, 1982).
\item[{[\rm 26]}] S. Weinberg, in
{\it General Relativity}, Eds. S.W. Hawking and W. Israel
( Cambridge University press, Cambridge, 1979) 790.
\item[{[\rm 27]}] N.N. Bogoljubov and D.V. Shirkov, {\it Introduction
to the theory of quantized fields, 3rd ed.}
(Wiley-Interscience, 1980).
\item[{[\rm 28]}] G. t'Hooft and M. Veltman, Ann. Inst. Henri
Poincare {\bf XX}, 69 (1974).
\item[{[\rm 29]}] M.H. Goroff and
A. Sagnotti, Nucl. Phys. B {\bf 266}, (1986) 709.
\item[{[\rm 30]}] P.van Nieuwenhuizen, Phys. Rep. {\bf 68C}, 189
(1981).
\item[{[\rm 31]}] M.B. Green, J.H. Schwarz and E. Witten,
{\it Superstring Theory, V.1,2} (Cambridge University Press,
Cambridge, 1987).
\item[{[\rm 32]}] L. Brink and M. Henneaux,
{\it Principles of String Theory} (Plenum Press, New York and London,
1988).
\item[{[\rm 33]}] S. Deser, in {\it Proceedings of the Conference
on Gauge Theories and Modern Field Theory}, Eds. R. Arnowitt, P.
Nath (MIT Press, Cambridge Massachusetts, 1975) 229.
\item[{[\rm 34]}] S. Weinberg, in {\it Proceedings of the XVIII
International conference on High Energy Physics}, Ed.J.R. Smith,
(Rutherford Laboratory, Chilton, Didcot, Oxfordshire, 1974) III-59.
\item[{[\rm 35]}] K.S. Stelle, Phys. Rev. D {\bf 16}, 953 (1977).
\item[{[\rm 36]}] D.I. Blokhintsev, A.V. Efremov and D.V. Shirkov,
Izv. VUZov, Fiz. (USSR) No 12, 23 (1974).
\item[{[\rm 37]}] D.I. Kazakov, Teor. Mat. Fiz. (USSR) {\bf 75},
 157 (1988).
\item[{[\rm 38]}] E.C.G. Stueckelberg and A. Petermann,
Helv. Phys. Acta {\bf 26}, 499 (1953).
\item[{[\rm 39]}] M. Gell-Mann and F.E. Low, Phys. Rev. {\bf 95},
 1300 (1954).
\item[{[\rm 40]}] A.A. Vladimirov and D.V. Shirkov,
Usp. Fiz. Nauk (USSR) {\bf 129}, 407 (1979).
\item[{[\rm 41]}] J.C. Collins, {\it Renormalization} (Cambridge
University press, Cambridge, 1984).
\item[{[\rm 42]}] Yu.V. Gryzov, A.Yu. Kamenshchik and I.P. Karmazin,
Izv. Vuzov, Fiz. (Sov.Phys.Journal) {\bf 35}, 121 (1992).
\item[{[\rm 43]}] E.S. Fradkin and G.A. Vilkovisky, On
renormalization of quantum field theory in curved spacetime,
Institute for Theoretical Physics preprint, University of
Berne (October, 1976).
\item[{[\rm 44]}] A.O. Barvinsky and G.A. Vilkovisky,
Phys. Rep. {\bf 119}, 1 (1985).
\item[{[\rm 45]}] B.S.DeWitt, Phys.Rev. {\bf 162}, 1195;
1239 (1967).
\item[{[\rm 46]}] B.S. DeWitt, {\it Dynamical Theory of Groups and
Fields} (Gordon and Breach, New York, 1965).
\item[{[\rm 47]}] S. Deser and P.van Nieuwenhuizen,
Phys. Rev. D {\bf 10}, 401 (1974),\\
S.Deser, Phys.Lett. B, {\bf 134}, 419 (1984).
\item[{[\rm 48]}] Don N.Page, J.Math..Phys. {\bf 32}, 3427 (1991).
\item[{[\rm 49]}] G.V.Bicknell, J. Phys. A {\bf 7}, 1061 (1974).
\item[{[\rm 50]}] B. Whitt, Phys. Lett. B {\bf 145}, 176 (1984).
\item[{[\rm 51]}] J.J. Halliwell, Phys. Lett. B {\bf 185},
341 (1987).
\item[{[\rm 52]}] A.B. Burd and J.D. Barrow,
Nucl. Phys. B {\bf 308}, 929 (1988).
\item[{[\rm 53]}] J.D. Barrow, Nucl. Phys. B, {\bf 296} 697 (1988).
\item[{[\rm 54]}] L.J. Garay, J.J. Halliwell and G.A. Mena Marugan,
Phys. Rev. D {\bf 43}, 2572 (1991).
\item[{[\rm 55]}] G. Denardo and
E. Spallucci, Nuovo Cim. A {\bf 68}, 177 (1982); {\bf 71}, 397
(1982), Lett.
Nuovo Cim. {\bf 34} 284 (1982).
\item[{[\rm 56]}] B.L. Nelson and P. Panangaden,
Phys. Rev. D {\bf 25}, 1019 (1982).
\item[{[\rm 57]}] I.L.Buchbinder and S.D.Odintsov, Izv. Vuzov,
Fiz.(Sov.Phys.Journal), {\bf N 12}, 108 (1983); Sov.J.Nucl.Phys. {\bf
40}, 1338
(1984).
\item[{[\rm 58]}] L. Parker and
D.J. Toms, Phys. Rev. Lett. {\bf 52}, 1269 (1984); Phys. Rev. D {\bf
29},
1584 (1984).
\item[{[\rm 59]}] L. Parker and D.J. Toms, Gen. Rel. Grav.
{\bf 17}, 167 (1985).
\item[{[\rm 60]}] I.L. Buchbinder,
Fortschr. Phys. {\bf 34}, 605 (1986).
\item[{[\rm 61]}] J. Guven, Phys. Rev. D {\bf 35}, 2378 (1987).
\item[{[\rm 62]}] I.L. Buchbinder, S.D. Odintsov and I.L. Shapiro,
Riv. Nuov. Cim. {\bf 12}, No 10 (1989).
\item[{[\rm 63]}] I.L. Buchbinder, S.D. Odintsov and I.L. Shapiro,
{\it
Effective Action in Quantum Gravity} (IOP, Bristol, 1992).
\item[{[\rm 64]}] G. t'Hooft, Nucl. Phys. B {\bf 61}, 455 (1973).
\item[{[\rm 65]}] G. t'Hooft, Nucl. Phys. B {\bf 62}, 444 (1973).
\item[{[\rm 66]}] I.F. Ginzburg, Docl. Acad. Nauk (USSR) {\bf 110},
 535 (1956).
\item[{[\rm 67]}] I.F. Ginzburg and D.V. Shirkov,
ZhETF (USSR) {\bf 49}, 335 (1965).
\item[{[\rm 68]}] D.V. Shirkov,
 Nucl. Phys. B {\bf 62}, 194 (1973).
\item[{[\rm 69]}] S. Coleman and
 E. Weinberg, Phys. Rev. D {\bf 7}, 1888 (1973).
\item[{[\rm 70]}] E.Mottola, Four dimensional quantum gravity: the
covariant
path integral and quantization of the conformal factor, Lectures
given at First
Iberian Conference on Gravity, Evora, Portugal, September 20-26,
1992.
\item[{[\rm 71]}] I.Antoniadis, P.O.Mazur and E.Mottola, Conformal
symmetry and
central charges in 4 dimensions, preprint LA-UR-1483, 1993, submitted
to
Nucl.Phys.B.
\end{description}
\end{document}